\documentclass{aastex631}
\usepackage{CJK}
\graphicspath{{./}{figures/}}
\begin{document}
\begin{CJK*}{UTF8}{gbsn}

\title{Stellar Chromospheric Activity Database of Solar-like Stars Based on the LAMOST Low-Resolution Spectroscopic Survey}

\correspondingauthor{Jun Zhang; Han He}
\email{zjun@ahu.edu.cn; hehan@nao.cas.cn}

\author{Weitao Zhang}
\affiliation{School of Physics and optoelectronics engineering, Anhui University, Hefei 230601, China} 

\author{Jun Zhang}
\affiliation{School of Physics and optoelectronics engineering, Anhui University, Hefei 230601, China}

\author{Han He}
\affiliation{CAS Key Laboratory of Solar Activity, National Astronomical Observatories, Chinese Academy of Sciences, Beijing 100101, China}
\affiliation{University of Chinese Academy of Sciences, Beijing 100049, China}

\author{Zhiping Song}
\affiliation{School of Physics and optoelectronics engineering, Anhui University, Hefei 230601, China}

\author{Ali Luo}
\affiliation{CAS Key Laboratory of Optical Astronomy, National Astronomical Observatories, Chinese Academy of Sciences, Beijing 100101, China}
\affiliation{University of Chinese Academy of Sciences, Beijing 100049, China}

\author{Haotong Zhang}
\affiliation{CAS Key Laboratory of Optical Astronomy, National Astronomical Observatories, Chinese Academy of Sciences, Beijing 100101, China}
\affiliation{University of Chinese Academy of Sciences, Beijing 100049, China}

\begin{abstract}
A stellar chromospheric activity database of solar-like stars is constructed based on the Large Sky Area Multi-Object Fiber Spectroscopic Telescope (LAMOST) Low-Resolution Spectroscopic Survey (LRS). 
The database contains spectral bandpass fluxes and indexes of Ca II H\&K lines derived from 1,330,654 high-quality LRS spectra of solar-like stars.
We measure the mean fluxes at line cores of the Ca II H\&K lines using a {1\,\AA} rectangular bandpass as well as a {1.09\,\AA} full width at half maximum (FWHM) triangular bandpass, and the mean fluxes of two {20\,\AA} pseudo-continuum bands on the two sides of the lines.
Three activity indexes, $S_{\rm rec}$ based on the {1\,\AA} rectangular bandpass, and $S_{\rm tri}$ and $S_L$ based on the {1.09\,\AA} FWHM triangular bandpass, are evaluated from the measured fluxes to quantitatively indicate the chromospheric activity level.
The uncertainties of all the obtained parameters are estimated.
We also produce spectrum diagrams of Ca II H\&K lines for all the spectra in the database.
The entity of the database is composed of a catalog of spectral sample and activity parameters, and a library of spectrum diagrams.
Statistics reveal that the solar-like stars with high level of chromospheric activity ($S_{\rm rec}>0.6$) tend to appear in the parameter range of $T_{\rm eff}\text{ (effective temperature)}<5500\,{\rm K}$, $4.3<\log\,g\text{ (surface gravity)}<4.6$, and $-0.2<[{\rm Fe/H}]\text{ (metallicity)}<0.3$.
This database with more than one million high-quality LAMOST LRS spectra of Ca II H\&K lines and basal chromospheric activity parameters can be further used for investigating activity characteristics of solar-like stars and solar-stellar connection.
\end{abstract}


\section{Introduction}\label{sec:introduction}

With detailed observations of solar activity for several centuries, many features and phenomena of the Sun, such as sunspots, plages, flares, etc., have been discovered and thoroughly studied. 
These features and phenomena are the manifestations of magnetic field activity on the Sun \citep{1908ApJ....28..315H}.
Observations for solar-like stars \citep{1996A&ARv...7..243C} revealed that magnetic activity is also common on other stars,
and the connection between stellar activity and solar activity (i.e., solar-stellar connection) has become a topic of wide interest \citep{1996ASPC..109....3N}.
\citet{2017SCPMA..60a9601C} collected various stellar activity data and explored the extrapolation of solar dynamo models for explaining magnetic activity of solar-like stars. 
The knowledge of activity of solar-like stars in turn is very helpful for understanding the activity status of the Sun \citep{2007LRSP....4....3G}. 
According to the classification by \citet{2021A&A...646A..77G}, the Sun is located in the high-variability region of the inactive main sequence star zone.
\citet{2020Sci...368..518R} illustrated that the Sun is less active compared with other solar-like stars.

Stellar activity is closely related with the rotation period (e.g., \citealt{1984ApJ...287..769N, 2016Natur.535..526W, 2020ApJS..247....9Z}) and age (e.g., \citealt{2008ApJ...687.1264M,2018A&A...619A..73L,2019ApJ...887...84Z}) of stars. In general, stellar activity level will decrease with increase in stellar rotation period or age. On the other hand, \citet{2012Natur.485..478M} found that the maximum energy of stellar flares is not correlated with stellar rotation period. 
Investigation on the relation between stellar activity cycle and rotation by \citet{2017A&A...603A..52R} 
reveals that the activity cycle period slightly increases for longer rotation period. 

As an important aspect of stellar activity, the chromospheric activity of solar-like stars has always been a popular research subject \citep{2008LRSP....5....2H}. 
A detailed review of stellar chromosphere modelling and spectroscopic diagnostics has been given by \citet{2017ARA&A..55..159L}. 
Stellar chromospheric activity of solar-like stars can be indicated by line core emissions of the Ca II H\&K lines in violet band of the visible spectrum (e.g., \citealt{1995ApJ...438..269B,2007AJ....133..862H}), the H$\alpha$ line in red band (e.g., \citealt{1998A&A...331..581D,2017ApJ...834...85N}), and the Ca II infrared triplet (Ca II IRT) lines (e.g., \citealt{1993ApJS...85..315S,2015PASJ...67...33N}), etc. The emission of Ca II H\&K lines of the Sun has long been known to have a strong correlation with the solar chromospheric activity (see a comprehensive review by \citealt{1970PASP...82..169L}).
With the discovery of the emissions of Ca II H\&K lines from other stars (e.g., \citealt{1913ApJ....38..292E}),
people began to explore whether it comes from the same mechanism as the solar activity and whether it has a long-term cyclic variation as the solar cycle.
\citet{1963ApJ...138..832W} at the Mount Wilson Observatory (MWO) investigated the relationship between intensity of stellar Ca II H\&K emission and stellar physical nature, and concluded that the chromospheric activity of main sequence stars decreases with age.
\citet{1978ApJ...226..379W} found the long-term cyclic variations of stellar Ca II H\&K fluxes similar to the solar cycle.
\citet{1995ApJ...438..269B} presented continuous Ca II H\&K emission records of stellar chromospheric activity for 111 stars, which came from a long-term observing program at MWO.
At that time, the MWO $S$ index was first introduced as a quantitative chromospheric activity indicator based on the Ca II H\&K lines \citep{1968ApJ...153..221W, 1978PASP...90..267V, 1991ApJS...76..383D, 1995ApJ...438..269B}, 
defined as the ratio between the flux of Ca II H\&K line cores and the flux of the reference bands on the violet and red sides of the lines multiplied by a scaling factor for calibration between different instruments.
Because of the high correlation between Ca II H\&K emission and stellar magnetic activity (e.g., \citealt{1987LNP...291...38S}), researchers prefer to characterize stellar magnetic activity through the indicators derived from Ca II H\&K lines which often involve $S$ index.

The $S$ index has been established as a fundamental parameter of stellar chromosphere activity. 
Based on $S$ index, by subtracting the photospheric contribution to Ca II H\&K flux, the true chromospheric emission of Ca II H\&K lines can be extracted as an index $R'_{\rm HK}$ \citep{1979ApJS...41...47L, 1984ApJ...279..763N}.
Considering the existence of the basal (lower-limit) flux of chromosphere \citep{1987A&A...172..111S} which is thought to be unrelated with magnetic activity, \citet{2013A&A...549A.117M} proposed a new index $R^{+}_{\rm HK}$ to reflect pure stellar chromospheric activity.

The Large Sky Area Multi-Object Fiber Spectroscopic Telescope (LAMOST, also named Guoshoujing Telescope) is a telescope dedicated for spectroscopic sky survey.
There are 4000 fibers within a diameter of 1.75 meters (corresponding to $5^\circ$ in the sky) at the focal surface \citep{2012RAA....12.1197C}.
The Low-Resolution Spectroscopic Survey (LRS) of LAMOST began in October 2011, with a spectral resolving power ($R=\lambda/\Delta\lambda$) of about 1800 and a wavelength coverage of 3700--9100\,{\AA} \citep{2012RAA....12..723Z}. 
The first year observation was for the pilot survey \citep{2012RAA....12.1243L, 2012RAA....12..723Z}, and the regular survey began in September 2012. 
LAMOST also conduct regular Medium-Resolution Spectroscopic Survey (MRS; $R \sim 7500$) since 2018 for wavelength bands of 4950--5350\,{\AA} and 6300--6800\,{\AA} \citep{2020arXiv200507210L}.
When completing seven years of sky survey (one year pilot survey and six years regular survey) in June 2018,
LAMOST became the first spectral sky survey project in the world to accumulate more than 10 million spectra.

\begin{figure}
    \begin{center}
        \includegraphics[width=0.98\textwidth]{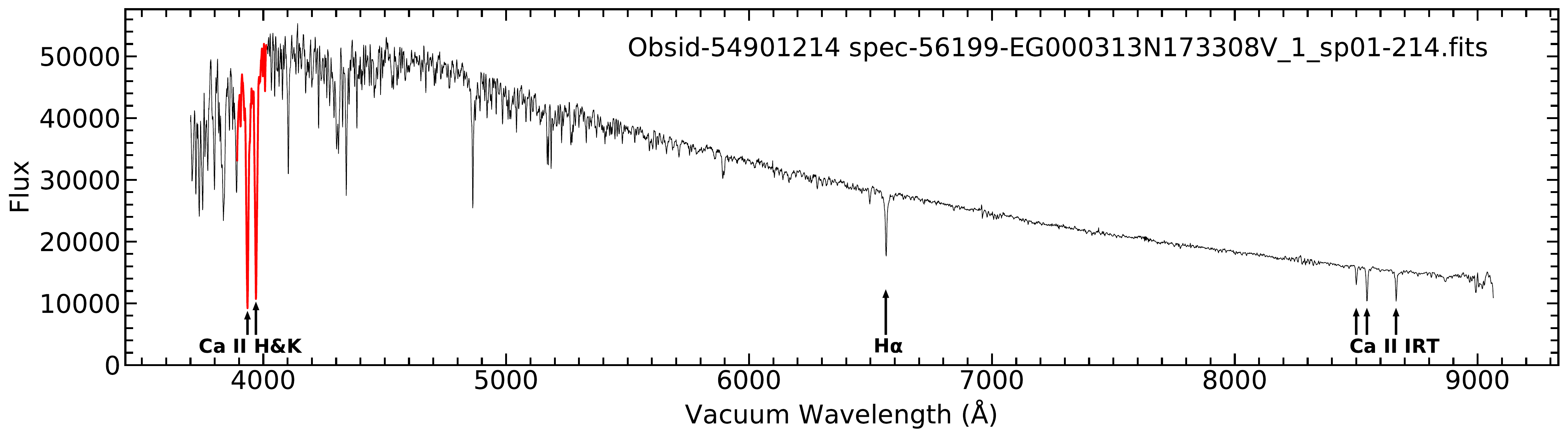}
    \end{center}
    \caption{An example of LAMOST LRS spectrum of solar-like stars. The spectral lines commonly used for analyzing stellar chromospheric activity (Ca II H\&K, H$\alpha$, and Ca II IRT) are labeled. The Ca II H\&K lines employed in this work are highlighted in red. The obsid (observation identifier) and FITS file name of the spectrum are shown in the plot for reference.
    \label{fig:LRS_spectrum_example}}
\end{figure}

The majority of the released data by LAMOST is LRS spectra. 
The huge data set is very beneficial for big-data analyses of stellar properties. There are millions of LRS spectra of solar-like stars that can be used for studying stellar chromospheric activity and solar-stellar connection.
Figure \ref{fig:LRS_spectrum_example} gives an example of LRS spectra of solar-like stars observed by LAMOST. 
As illustrated in Figure \ref{fig:LRS_spectrum_example}, the LRS spectrum contains several optical spectroscopic features that can indicate stellar chromospheric activity, in which the Ca II H\&K lines (highlighted in red) are the most commonly employed spectral lines.
The other lines (H$\alpha$ and Ca II IRT) can also be used (e.g., \citealt{2016A&A...594A..39F}), but those lines are relatively narrow and hence are not well-resolved in LRS spectra compared with Ca II H\&K lines.

The massive amount of LAMOST LRS spectral data provides a great opportunity for investigating overall chromospheric activity properties of solar-like stars based on the Ca II H\&K lines.
\citet{2016NatCo...711058K} selected 5,648 solar-like stars (including 48 superflare stars) from LAMOST LRS catalog and 
studied the relation between stellar chromospheric activity level (indicated by the Ca II H\&K $S$ index) and occurrence of superflares.
\citet{2020ApJS..247....9Z} calculated chromospheric $S$ index and $R^+_{\rm HK}$ index of 59,816 stars from the LRS spectra of the LAMOST-\emph{Kepler} observation project \citep{2015ApJS..220...19D, 2018ApJS..238...30Z, 2020RAA....20..167F} and investigated the relationship between stellar chromospheric activity and photospheric activity (indicated by the stellar light curves observed by \emph{Kepler} space telescope) of  F-, G-, and K-type stars and the dependence of the activities on stellar rotation.
\citet{2015RAA....15.1282Z} presented measurements of chromospheric $S$ index for 119,995 F, G and K stars by using the LRS spectra from the first data release of LAMOST.
\citet{2021ApJS..253...35T} found 7,454 solar-like stars with both light-curve observation by the Transiting Exoplanet Survey Satellite (TESS; \citealt{2015JATIS...1a4003R}) and LRS spectral observation by LAMOST and investigated the relations between the stellar chromospheric activity (measured by $S$ index), photometric variability, and flare activity of the stellar objects.

The previous works discussed above used a subset of the released LAMOST spectra. We believe that exploiting the full data set of LAMOST will further facilitate the relevant research \citep{2021RNAAS...5....6H}.
By utilizing the large volume of LRS spectra in LAMOST Data Release 7,
we constructed a stellar chromospheric activity database of solar-like stars based on the Ca II H\&K lines, which is elaborated in this paper.
We introduce the LAMOST Data Release 7 in Section \ref{sec:lamost-dr} and explain the criteria for selecting high-quality LRS spectra of solar-like stars in Section \ref{sec:lrs-spectra-selection}.
The data processing workflow for the selected LRS spectra and the derivation of stellar chromospheric activity measures and indexes are described in detail in Section \ref{sec:data-processing}.
The components of the stellar chromospheric activity database of solar-like stars are elucidated in Section \ref{sec:database}.
We discuss the results obtained from the database and perform a statistical analysis of the stellar chromospheric activity indexes in Section \ref{sec:results}.
In Section \ref{sec:conclusion}, we summarize this work and prospect the further researches based on the database.

\section{Data Release of LAMOST} \label{sec:lamost-dr}

The annual observation of the LAMOST sky survey begins in September of each year and ends in June of the next year. 
The summer season (from July to August) is for instrument maintenance. 
The acquired data by LAMOST are also released in yearly increments.
Each data release (DR) consists of the data files of one-dimensional spectra (in FITS format) and the catalog files (in both FITS and CSV formats) of spectroscopic parameters \citep{2015RAA....15.1095L}. 
A new DR contains all the data collected in the corresponding observing year as well as in the previous years.

The LAMOST Data Release 7 (DR7) is opened to the public in September 2021, which contains the spectral data observed from October 2011 to June 2019.
In this work, we utilize the LRS spectra in LAMOST DR7 v2.0\footnote{\url{http://www.lamost.org/dr7/v2.0/}} to construct the stellar chromospheric activity database of solar-like stars. 
As demonstrated in Figure \ref{fig:LRS_spectrum_example}, the flux of LRS spectra is calibrated, and the vacuum wavelength is adopted in LAMOST data.
Most telluric lines in red band of LRS spectra have been removed \citep{2015RAA....15.1095L}.

Basic information of the LRS spectra in LAMOST DR7 v2.0 is shown in Table \ref{tab:lamost-dr7-lrs-info}. 
The spectra of solar-like stars investigated in this work are taken from the {\tt\string LAMOST LRS Stellar Parameter Catalog of A, F, G and K Stars} (hereafter referred to as {\tt\string LAMOST LRS AFGK Catalog}, for short), which consists of 48 spectroscopic parameters, 
such as observation identifier (obsid), sky coordinates, signal-to-noise ratio (SNR), magnitude, and so on.
Several important stellar parameters, including effective temperature ($T_{\rm eff}$), surface gravity ($\log\,g$), metallicity ([Fe/H]), and radial velocity ($V_r$), are also provided by the catalog.
The four stellar parameters are obtained by the LAMOST Stellar Parameter Pipeline (LASP), in which all the parameters are determined simultaneously by minimizing the squared difference between the observed spectra and the model spectra \citep{2011RAA....11..924W, 2015RAA....15.1095L}.
The targets in LAMOST DR7 have been cross-matched with the Gaia DR2 catalog \citep{2018A&A...616A...1G}, and their Gaia source identifiers and $G$ magnitudes are included in the LAMOST DR7 catalogs.

\begin{deluxetable}{cc}
\tablecaption{Basic information of the LRS spectra in LAMOST DR7 v2.0. \label{tab:lamost-dr7-lrs-info}}
\tablehead{
\colhead{Classification of LRS spectra} & \colhead{Number of LRS Spectra}
}
\startdata
Total Released Spectra & 10,431,197 \\
Stellar Spectra & 9,846,793 \\
Stellar Spectra with ${\rm SNR}_g\:\text{or}\:{\rm SNR}_i > 10$  & 8,912,384 \\
Stellar Spectra with AFGK Stellar Parameters & 6,179,327 \\
High-Quality Spectra of Solar-like Stars* & 1,330,654
\enddata
\tablecomments{The first four rows are taken from LAMOST DR7 v2.0. 
${\rm SNR}_g$ and ${\rm SNR}_i$ mean signal-to-noise ratio in $g$ band and $i$ band (see Section \ref{sec:lrs-spectra-selection} for an explanation), respectively.
The number of high-quality LRS spectra of solar-like stars (labeled with a `\textasteriskcentered' symbol) is obtained in this work (see Section \ref{sec:lrs-spectra-selection} for details).}
\end{deluxetable}

\section{Selection of High-Quality LRS Spectra of Solar-like Stars}\label{sec:lrs-spectra-selection}
High-quality LAMOST LRS spectra of solar-like stars \citep{1996A&ARv...7..243C} are expected in this work. 
The Sun is a main-sequence star with spectral class of G2, $T_{\rm eff}$ of about 5800 K, and $\log\,g$ of about 4.4 ($g$ in unit of $\mathrm{cm}\cdot\mathrm{s}^{-2}$).
There are various definitions for solar-like stars in the literature with different ranges of stellar parameters around the values of the Sun (e.g., \citealt{2000ApJ...529.1026S, 2012Natur.485..478M, 2013ApJS..209....5S, 2015RAA....15.1282Z, 2020Sci...368..518R, 2020ApJS..247....9Z, 2020ApJ...894L..11Z}). 
In this work, we consider the concept of solar-like stars from the viewpoint of stellar chromospheric activity and adopt a broader range of spectral class than G-type, since Ca II H\&K lines are also prominent in spectra of late F- and early K-type stars (e.g., \citealt{2017ApJS..230...16K}).

We select high-quality LRS spectra of solar-like stars from the {\tt\string LAMOST LRS AFGK Catalog} of LAMOST DR7 v2.0, which contains 6,179,327 LRS spectra (see Table \ref{tab:lamost-dr7-lrs-info}) with determined stellar parameters ($T_{\rm eff}$, $\log\,g$, [Fe/H], and $V_r$) by the LASP. 
The criteria for the spectral data selection involve five aspects: SNR condition, $T_{\rm eff}$ range, [Fe/H] range, main-sequence star condition, and data completeness in Ca II H\&K band. The SNR and data completeness criteria are for the high-quality data, and the $T_{\rm eff}$, [Fe/H], and main-sequence star criteria are for the sample of solar-like stars. These criteria are described as follows.

\begin{enumerate}
    \item 
One major indicator of the quality of a spectrum is the SNR. 
LAMOST catalogs provide SNR parameters for LRS spectra in five color bands, that is, ultraviolet, green, red, near infrared, and infrared bands, which are abbreviated as $u$, $g$, $r$, $i$, and $z$, respectively.\footnote{The $ugriz$ color bands have been adopted by the Sloan Digital Sky Survey (SDSS) \citep{2002AJ....123..485S}.}
In LAMOST catalogs, the value of SNR is in the range from 0 to 1000.
Higher SNR generally means higher quality of the spectral data, and hence smaller uncertainties of the spectral fluxes, determined stellar parameters, and derived stellar chromospheric activity parameters.
In this work, we utilize the SNR parameters of LRS spectra in the $g$ band and $r$ band\footnote{The wavelength ranges of the $g$ band and $r$ band are 3620--5620\,{\AA} and 5380--7040\,{\AA}, respectively \citep{2010AJ....139.1628D}.} (denoted by ${\rm SNR}_g$ and ${\rm SNR}_r$, respectively), and adopt the SNR condition for high-quality LRS spectra as ${\rm SNR}_g \ge 50.00$ and ${\rm SNR}_r \ge 71.43$.
This criterion is a compromise between a smaller uncertainty of spectral fluxes/stellar parameters/activity parameters and a larger volume of spectral sample.
The $g$-band threshold (${\rm SNR}_g \ge 50.00$) is the primary condition;
the $r$-band threshold (${\rm SNR}_r \ge 71.43$) is determined from the $g$-band condition in consideration that ${\rm SNR}_r/{\rm SNR}_g \sim 10/7$ for the spectra with $T_\mathrm{eff}$ in the range of solar-like stars (see criterion 2).
Figure \ref{fig:LRS_snrg-snrr} shows the scatter plot of ${\rm SNR}_g$ vs. ${\rm SNR}_r$, in which the ratio of ${\rm SNR}_r/{\rm SNR}_g=10/7$ and the SNR thresholds for high-quality spectra are illustrated.
It should be noted that although the Ca II H\&K lines employed in this work are in the $g$ band, the whole LRS spectrum is used by the LASP to determine the stellar parameters \citep{2015RAA....15.1095L}. 
In addition to the $g$-band SNR condition, the $r$-band SNR condition is included to ensure a smaller uncertainty of stellar parameters.

\begin{figure}
	\begin{center}
	    \includegraphics[width=0.50\textwidth]{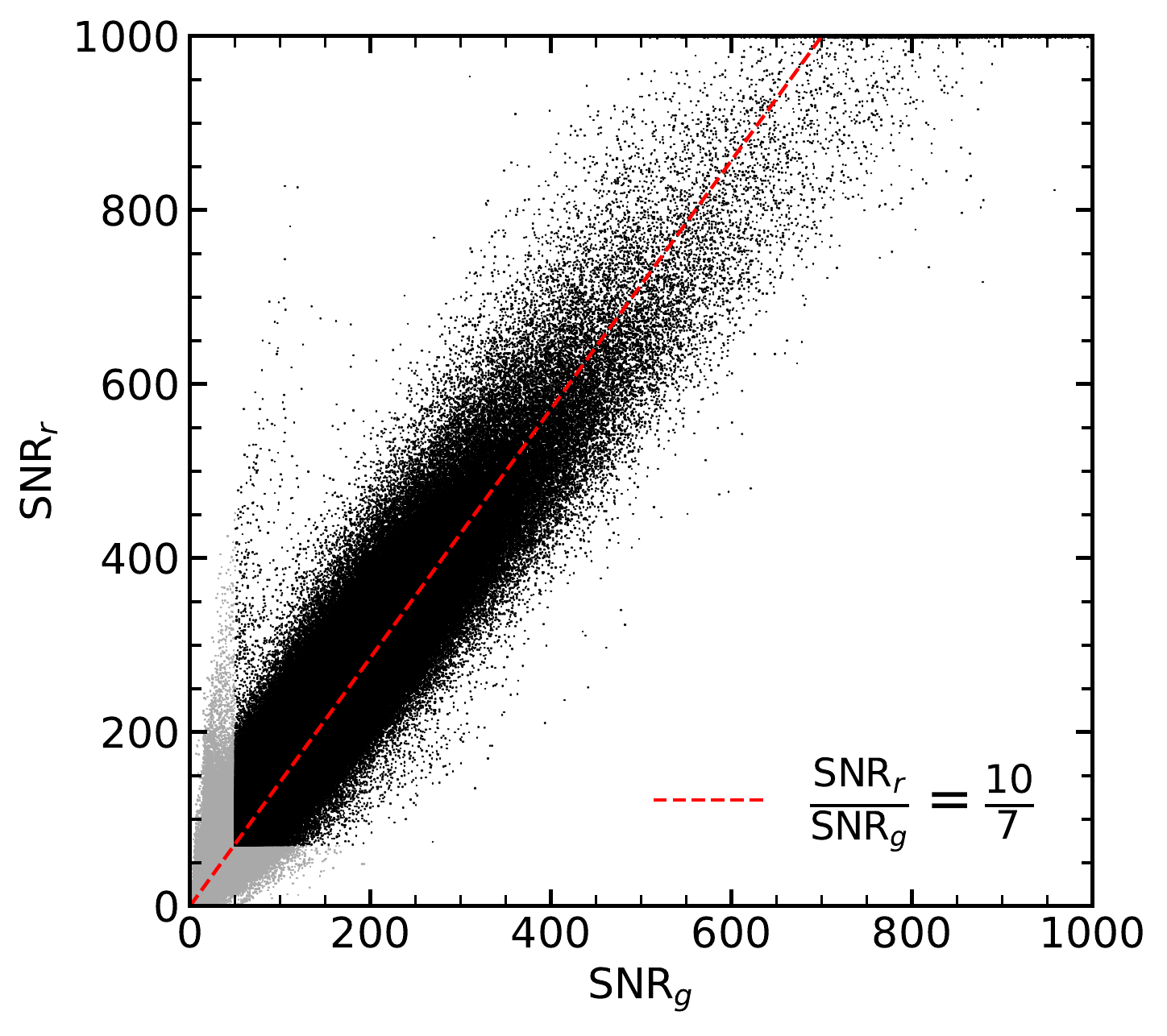}
	\end{center}
    \caption{Scatter plot of ${\rm SNR}_g$ vs. ${\rm SNR}_r$ for the spectra in the {\tt\string LAMOST LRS AFGK Catalog} with $T_\mathrm{eff}$ in the range of 4800--6800\,K.
    The black dots represent the high-quality spectra with ${\rm SNR}_g \ge 50.00$ and ${\rm SNR}_r \ge 71.43$. 
    The gray dots represent the spectra with ${\rm SNR}_g$ or ${\rm SNR}_r$ below the thresholds. 
    The dashed line indicates the ratio of ${\rm SNR}_r/{\rm SNR}_g = 10/7$. 
    \label{fig:LRS_snrg-snrr}}
\end{figure}

    \item 
The range of $T_\mathrm{eff}$ for solar-like stars is adopted as $4800\,{\rm K} \le T_\mathrm{eff} \le 6800\,{\rm K}$, which is around the effective temperature of the Sun ($\sim 5800$\,K) with fluctuation of $\pm 1000$\,K. The main-sequence stars with $T_\mathrm{eff}$ in this range presents prominent Ca II H\&K lines in their spectra (e.g., \citealt{2017ApJS..230...16K}) and hence are considered as candidates of solar-like stars; they belong to late F-type, G-type, and early K-type stars.

    \item
The range of [Fe/H] for solar-like stars is adopted as $-1.0<\text{[Fe/H]}<1.0$, which is around the metallicity of the Sun ($\text{[Fe/H]}=0.0$). The stars with [Fe/H] less than $-1.0$ (metal-poor stars) might have a different origin from the Sun (e.g., \citealt{2015ARA&A..53..631F}) and hence are not considered to be solar-like.

    \item
In the $T_\mathrm{eff}$ - $\log\,g$ diagram of stellar objects, main-sequence stars distribute in the horizontal branch and giant stars are in the erect branch. Base on the $T_\mathrm{eff}$ - $\log\,g$ diagram of the spectra in the {\tt\string LAMOST LRS AFGK Catalog}, as shown in Figure \ref{fig:LRS_teff-logg}, we adopt the following empirical formula to distinguish the spectra of main-sequence stars in the $T_{\rm eff}$ range of 4800--6800\,K:
\begin{equation}\label{eq:main-sequence_condition}
    \log\,g \ge 5.98-0.00035 \times T_{\rm eff} \qquad \textrm{for $4800\,{\rm K} \le T_{\rm eff} \le 6800$\,K}.
\end{equation}
The separation line between the giant and main-sequence samples defined by Equation (\ref{eq:main-sequence_condition}) is shown as a black solid line ($\log\,g = 5.98-0.00035 \times T_{\rm eff}$) in Figure \ref{fig:LRS_teff-logg}.
The two endpoints of the line, ($T_{\rm eff}=6800\,{\rm K}$, $\log\,g=3.6$) and ($T_{\rm eff}=4800\,{\rm K}$, $\log\,g=4.3$), are determined empirically by visual inspecting the distribution of the LRS samples in the $T_\mathrm{eff}$ - $\log\,g$ diagram.
In Figure \ref{fig:LRS_teff-logg}, the sample of main-sequence stars (beneath the black solid line, as according to Equation (\ref{eq:main-sequence_condition})) with $T_\mathrm{eff}$ in the range of 4800--6800 K (criterion 2) and SNR above the thresholds (criterion 1) is shown in green; 
the sample of giant stars (above the black solid line) with SNR above the thresholds is shown in orange; 
the sample of other main-sequence stars ($T_{\rm eff}<4800$\,K or $T_{\rm eff}>6800$\,K) with SNR above the thresholds is shown in blue;
and the sample below the SNR thresholds is shown in gray.

\begin{figure}
    \begin{center}
    \includegraphics[width=0.79\textwidth]{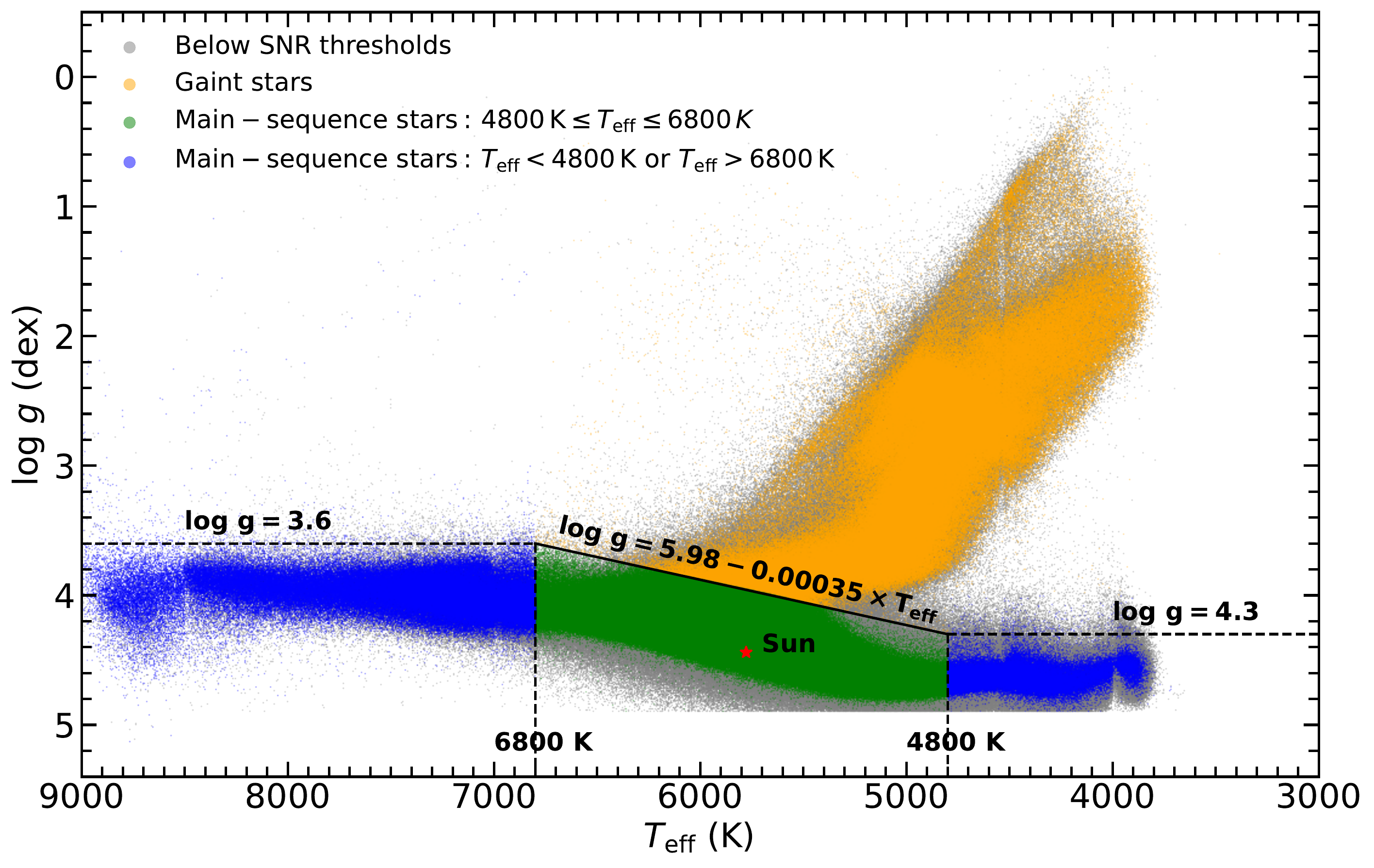}
    \end{center}
    \caption{$T_{\rm eff}$ - $\log\,g$ diagram of the spectra contained in the {\tt\string LAMOST LRS AFGK Catalog}. 
    The black solid line is defined by Equation (\ref{eq:main-sequence_condition}) and divides the samples of main-sequence stars (beneath the line) and giant stars (above the line). 
    The sample with SNR below the thresholds (see criterion 1 in Section \ref{sec:lrs-spectra-selection}) is shown in gray. 
    The sample with SNR above the thresholds is shown in color, in which 
    the sample of main-sequence stars with $4800\,{\rm K} \le T_{\rm eff} \le 6800\,{\rm K}$ (solar-like candidates) is shown in green, 
    the sample of other main-sequence stars ($T_{\rm eff}<4800$\,K or $T_{\rm eff}>6800$\,K) is in blue, and
    the sample of giant stars is in orange. 
    The position of solar $T_{\rm eff}$ and $\log\,g$ is indicated with a `$\star$' symbol.
    \label{fig:LRS_teff-logg}}
\end{figure}

    \item
In this work, we utilize the Ca II H\&K band to analyze stellar chromospheric activity. 
Some LRS spectral data in this band contain data points with zero or negative fluxes, which are not reliable according to the caveat of LAMOST and should be removed from the spectral data.
Those LRS spectra with incomplete data points in Ca II H\&K band are not used in the analysis.

\end{enumerate}

\begin{figure}
	\begin{center}
	\includegraphics[width=0.50\textwidth]{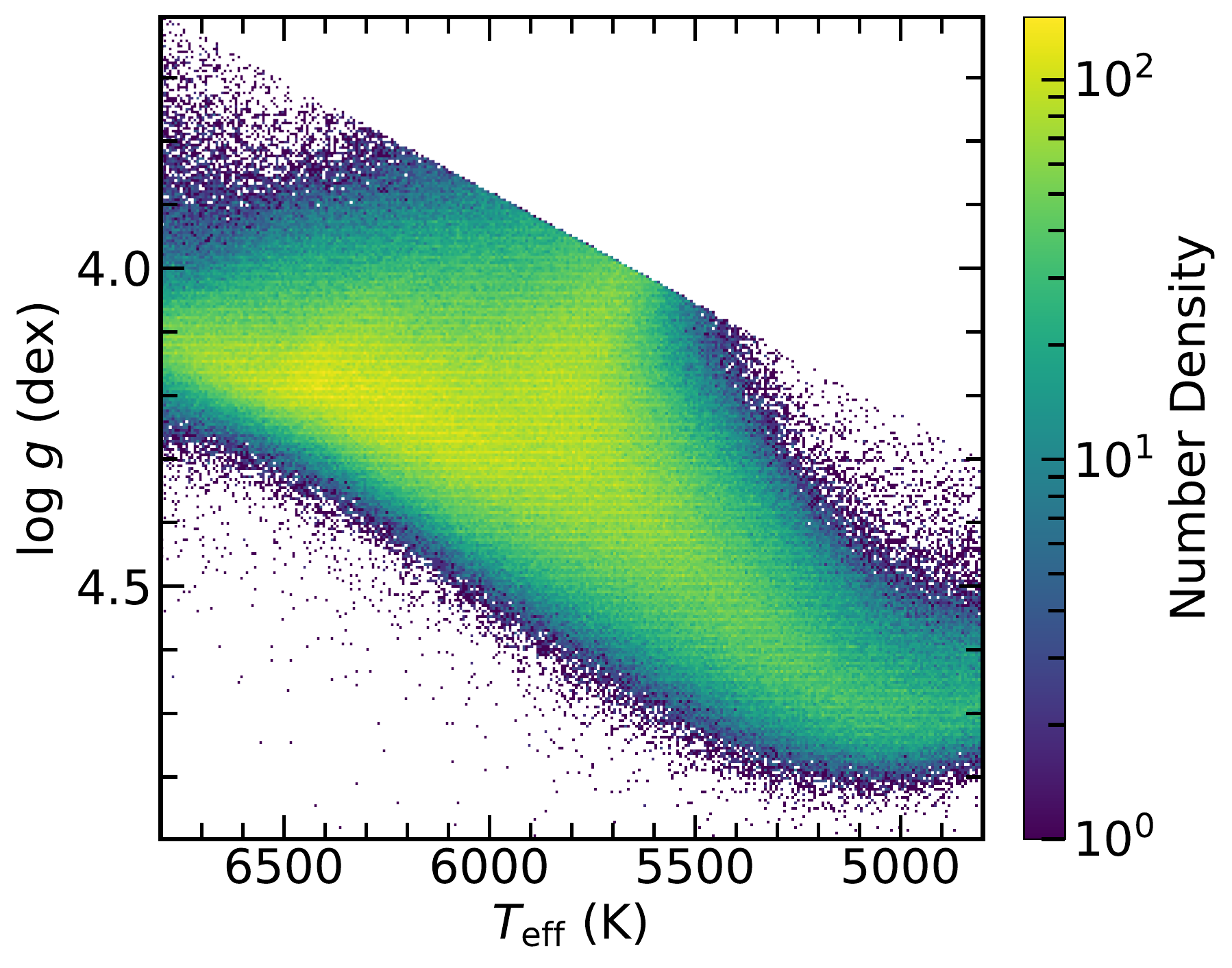}
	\end{center}
	\caption{Number density distribution of all the selected high-quality LAMOST LRS spectra of solar-like stars in the $T_{\rm eff}$ - $\log\,g$ parameter space.
	\label{fig:LRS_solar-like_teff-logg}}
\end{figure}

\begin{figure}
    \begin{center}
    \includegraphics[width=0.85\textwidth]{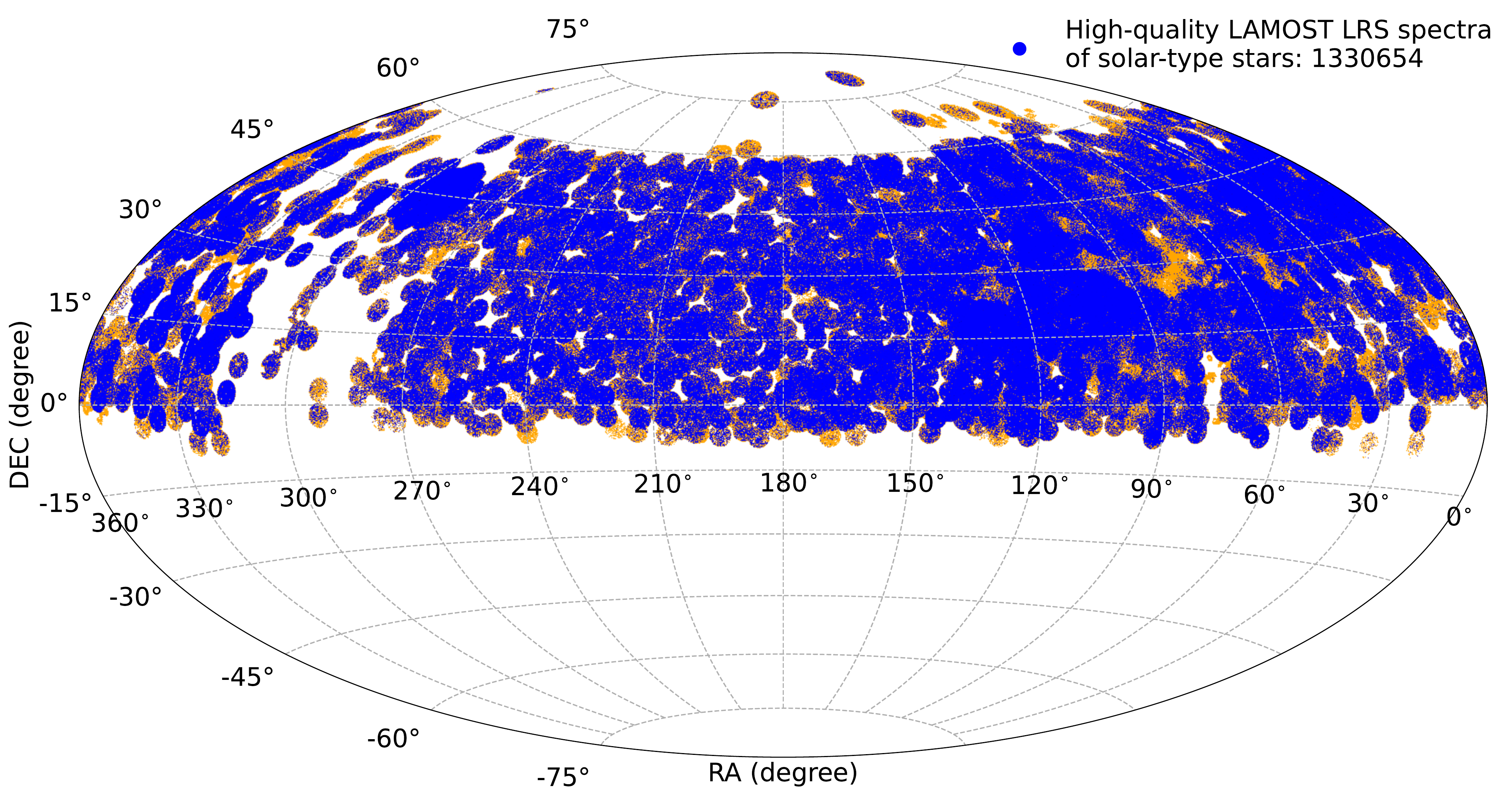}
    \end{center}
    \caption{Diagram illustrating sky coordinates of all the selected high-quality LAMOST LRS spectra of solar-like stars (blue dots). The original sample in the {\tt\string LAMOST LRS AFGK Catalog} is also depicted for reference (in orange and beneath the blue dots).
    \label{fig:LRS_solar-like_skymap}}
\end{figure}

By applying the first four criteria (SNR condition, $T_\mathrm{eff}$ range, [Fe/H] range, and main-sequence star condition) for the spectra in the {\tt\string LAMOST LRS AFGK Catalog}, we get a sample of 1,352,910 LRS spectra. 
By applying the fifth criterion (data completeness in Ca II H\&K band), 22,256 spectra in the sample are discarded.
We ultimately obtain 1,330,654 high-quality LAMOST LRS spectra of solar-like stars that are suitable for studying stellar chromospheric activity through the Ca II H\&K lines.
The number density distribution of these selected LRS spectra in the $T_\mathrm{eff}$ - $\log\,g$ parameter space is shown in Figure \ref{fig:LRS_solar-like_teff-logg}. 
The sky coordinates of these selected spectra are illustrated in Figure \ref{fig:LRS_solar-like_skymap}.

\begin{figure}
    \epsscale{0.85}
	\plotone{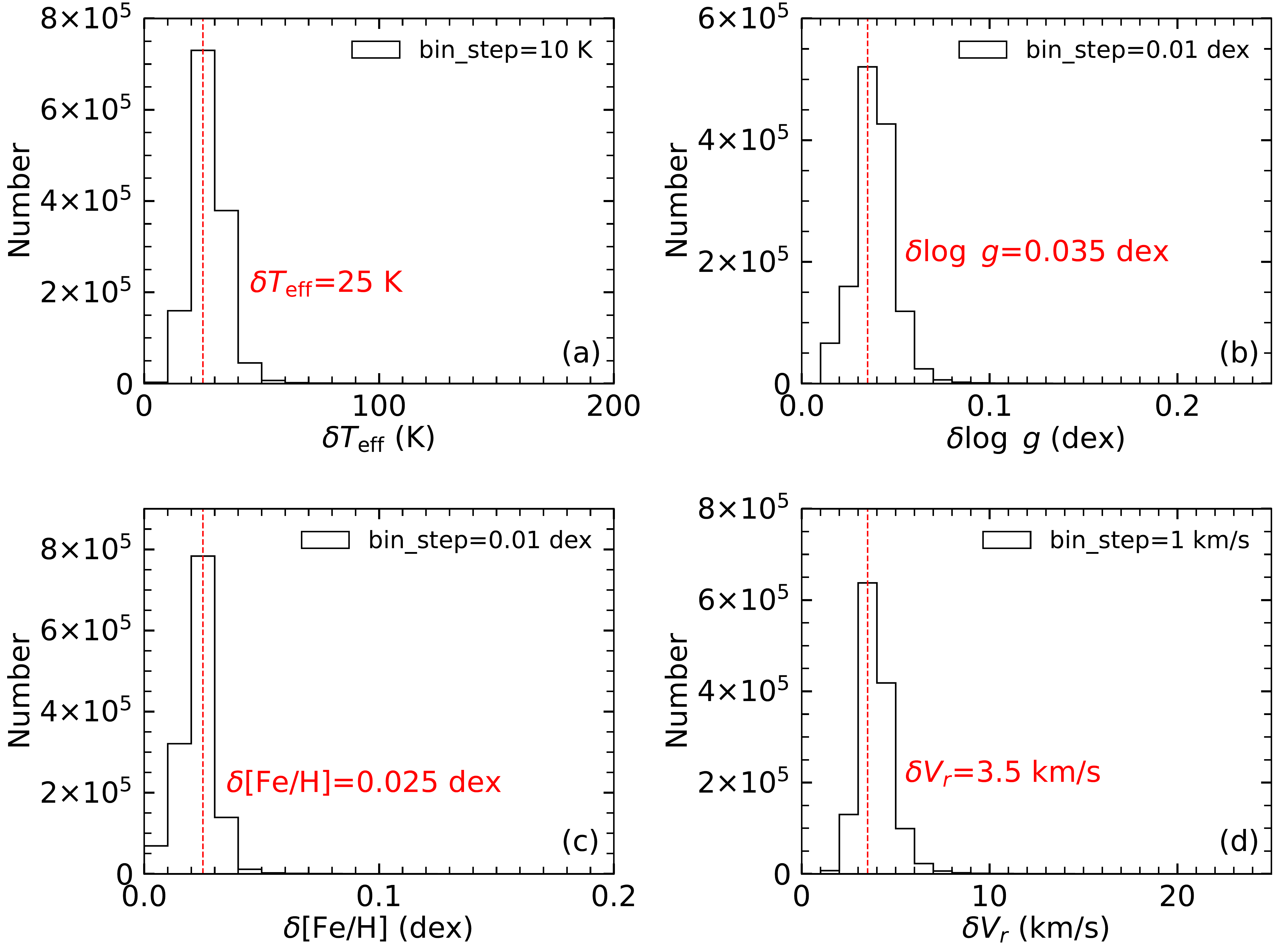}
	\caption{Histograms of the uncertainty values of (a) $T_{\rm eff}$, (b) $\log\,g$, (c) [Fe/H], and (d) $V_r$ for all the selected high-quality LAMOST LRS spectra of solar-like stars. The vertical dashed line in each plot indicates the peak position of the uncertainty distribution, and the value of the peak position is labeled.}
    \label{fig:teff-logg-feh-rv_err}
\end{figure}

In Figure \ref{fig:teff-logg-feh-rv_err}, we show the distributions of the uncertainty values of the four stellar parameters determined by the LASP (denoted by $\delta T_{\rm eff}$, $\delta \log\,g$, $\delta {\rm [Fe/H]}$, and $\delta V_r$, respectively) for all the selected high-quality LRS spectra of solar-like stars. 
It can be seen from Figure \ref{fig:teff-logg-feh-rv_err} that the peak positions of the uncertainty distributions of the four stellar parameters are $\delta T_{\rm eff} \sim 25\,{\rm K}$, $\delta \log\,g \sim 0.035\,{\rm dex}$, $\delta {\rm [Fe/H]} \sim 0.025\,{\rm dex}$, and $\delta V_r \sim 3.5\,{\rm km/s}$, respectively. 
The high-accurate stellar parameters of the selected LRS spectra of solar-like stars lay a good foundation for the subsequent data processing and analysis.

\section{Data Processing Workflow for the Selected LRS Spectra of Solar-like Stars}\label{sec:data-processing}

The purpose of data processing is to obtain emission fluxes of Ca II H\&K lines and chromospheric activity indexes for each of the selected LRS spectra of solar-like stars. 
The workflow for data processing has four steps \citep{2021RNAAS...5....6H}: 
(1) correction for wavelength shift introduced by radial velocity;
(2) measurement of emission fluxes of Ca II H\&K lines;
(3) evaluation of chromospheric activity indexes; and
(4) estimation of uncertainties of emission flux measures and activity indexes.
These steps are described in detail in the following subsections.

\subsection{Correction for Wavelength Shift Introduced by Radial Velocity} \label{sec:rv-correction}
The radial velocity of a stellar object causes wavelength shift in the observed spectrum.
This wavelength shift should be corrected to obtain the spectrum in rest frame before measuring the emission fluxes of Ca II H\&K lines.
The radial velocity values of the selected LRS spectra of solar-like stars have been determined by the LASP and are provided by the {\tt\string LAMOST LRS AFGK Catalog}.

The relation between the radial velocity $V_r$ and the wavelength shift of a spectrum can be expressed as
\begin{equation} \label{eq:lambda-rv}
	\frac{\lambda-\lambda_0}{\lambda_0} = \frac{V_r}{c},
\end{equation}
where $c$ is the speed of light, $\lambda$ is the wavelength value of the observed spectrum, $\lambda_0$ is the wavelength value in rest frame, and $\lambda-\lambda_0$ is the wavelength shift introduced by radial velocity.
Then, the desired wavelength value in rest frame, $\lambda_0$, can be calculated from the observed wavelength value $\lambda$ and radial velocity $V_r$ by the following equation:
\begin{equation} \label{eq:lambda_0}
	\lambda_0 = \frac{\lambda}{\displaystyle 1+\frac{V_r}{c}}.
\end{equation}
An example of wavelength shift correction for Ca II H\&K lines in LAMOST LRS spectra can be seen in Figure \ref{fig:spectrum_diagram_example}.

\subsection{Measurement of Emission Fluxes of Ca II H\&K lines} \label{sec:flux_measurements}

The first long-term program of measuring emission fluxes of stellar Ca II H\&K lines for a bulk of stars started in 1966 at MWO \citep{1968ApJ...153..221W} using a photoelectric scanner called HKP-1.
When a new photoelectric spectrometer referred to as HKP-2 was constructed at MWO, \citet{1978PASP...90..267V} formally defined the $H$, $K$, $R$ and $V$ channels of Ca II H\&K lines and began to routinely measure stellar emissioin fluxes in the four bands.
The $H$ and $K$ bands are two 1.09\,{\AA} full width at half maximum (FWHM) triangular bandpasses at line cores of Ca II H and K lines (center wavelengths in air being 3968.47\,{\AA} and 3933.66\,{\AA}, respectively), 
and the $R$ and $V$ bands are two 20\,{\AA} rectangular bandpasses on the red and violet sides of the Ca II H\&K lines (wavelength ranges in air being 3991.07--4011.07\,{\AA} and 3891.07--3911.07\,{\AA}, respectively). 
The $R$ and $V$ bands provide reference fluxes of pseudo-continuum for evaluation of $S$ index \citep{1978PASP...90..267V, 1991ApJS...76..383D}.

The aforementioned $H$, $K$, $R$, and $V$ bands have become the standard for characterizing emissions of Ca II H\&K lines and been widely used for assessing stellar chromospheric activity in the literature (e.g., \citealt{2009AJ....138..312H, 2010ApJ...725..875I, 2013A&A...549A.117M, 2016A&A...589A..27B, 2016A&A...596A..31S, 2018A&A...616A.108B, 2019MNRAS.485.5096K, 2020AJ....160..269M, 2020ApJS..247....9Z, 2021A&A...646A..77G, 2021ApJ...914...21S}). 
Meanwhile, researchers also look for alternative definitions of the bands to quantify the emissions of Ca II H\&K lines for specific facilities and data sets.
For example, \citet{2013AJ....145..140Z} defined their $H$ and $K$ bands as 2\,{\AA} rectangular bandpasses for the spectral data of SDSS;
\citet{2016A&A...591A..89M} adopted 1\,{\AA} rectangular $H$ and $K$ bands for the spectral data of the TIGRE telescope.
Both \citet{2013AJ....145..140Z} and \citet{2016A&A...591A..89M} kept the definitions of $R$ and $V$ bands by MWO. 

In this work, we adopt the classical definitions of $R$ and $V$ bands with 20\,{\AA} rectangular bandpasses, the classical definitions of $H$ and $K$ bands with 1.09\,{\AA} FWHM triangular bandpasses, and the alternative definitions of $H$ and $K$ bands with 1\,{\AA} rectangular bandpasses for the LAMOST LRS spectra. 
Six emission flux measures are introduced to evaluate the mean fluxes in the six bandpasses, which are tabulated in Table \ref{tab:CaIIHK-emission-flux-measures}. 
These measures are 
$\widetilde{R}$ and $\widetilde{V}$ for the mean fluxes in the 20\,{\AA} wide $R$ and $V$ reference bands,
$\widetilde{H}_{\rm tri}$ and $\widetilde{K}_{\rm tri}$ for the mean fluxes in the $H$ and $K$ bands with 1.09\,{\AA} FWHM triangular bandpasses, 
and $\widetilde{H}_{\rm rec}$ and $\widetilde{K}_{\rm rec}$ for the mean fluxes in the $H$ and $K$ bands with 1\,{\AA} rectangular bandpasses.
$\widetilde{H}_{\rm rec}$ and $\widetilde{K}_{\rm rec}$ are measured for Ca II H\&K lines in addition to $\widetilde{H}_{\rm tri}$ and $\widetilde{K}_{\rm tri}$ because the physical meaning of the fluxes in rectangular bandpass are more straightforward than the fluxes in triangular bandpass (see further discussion on the two types of bandpasses in Section \ref{sec:tri_vs_rec}).

The center wavelength values of the $R$, $V$, $H$, and $K$ bands are given in the rightmost two columns of Table \ref{tab:CaIIHK-emission-flux-measures}. 
The wavelength values in air are taken from \citet{1978PASP...90..267V},
and the wavelength values in vacuum are calculated from the air wavelengths by using the conversion formula given by \citet{1996ApOpt..35.1566C}.
In practical computation, the vacuum wavelengths are employed to derive the six emission flux measures from the LAMOST LRS spectra.
A diagram illustration of the vacuum wavelength ranges of the 20\,{\AA} wide $R$ and $V$ bands, the 1.09\,{\AA} FWHM triangular $H$ and $K$ bands, and the 1\,{\AA} rectangular $H$ and $K$ bands can be found in Figure \ref{fig:spectrum_diagram_example}.

We derive the values of the six emission flux measures ($\widetilde{R}$, $\widetilde{V}$, $\widetilde{H}_{\rm tri}$, $\widetilde{K}_{\rm tri}$, $\widetilde{H}_{\rm rec}$, and $\widetilde{K}_{\rm rec}$) of Ca II H\&K lines for all the selected LRS spectra of solar-like stars. 
To measure the mean flux in a bandpass for a LAMOST LRS spectrum, we first integrate the spectral fluxes in the bandpass, and then divide the integrated flux value by the wavelength width of the bandpass.
Since the LRS spectrum consists of discrete data points which are a bit sparse for the bandpass integration,
we obtain a denser distribution of data points in the bandpass via linear interpolation.
The wavelength steps after interpolation are 0.01\,{\AA} for $R$ and $V$ bands and 0.001\,{\AA} for $H$ and $K$ bands. 
Then, the mean flux value in a bandpass is calculated based on the interpolated spectral data.

\begin{deluxetable}{cclcc}
\tablecaption{Emission flux measures of Ca II H\&K lines. \label{tab:CaIIHK-emission-flux-measures}}
\tablehead{
\colhead{Measure} & \colhead{Column in} & \colhead{Description} & \multicolumn{2}{c}{Center Wavelength} \\
\cline{4-5}
 & \colhead{Database} & &  \colhead{(in Air)} & \colhead{(in Vacuum)}
}
\startdata
$\widetilde{R}$ & {\tt\string R\_mean} & Mean flux in $R$ band (20\,{\AA} rectangular bandpass) & 4001.07\,{\AA} & 4002.20\,{\AA} \\
$\widetilde{V}$ & {\tt\string V\_mean} & Mean flux in $V$ band (20\,{\AA} rectangular bandpass) & 3901.07\,{\AA} & 3902.17\,{\AA} \\
$\widetilde{H}_{\rm tri}$ & {\tt\string H\_mean\_tri} & Mean flux in $H$ band with 1.09\,{\AA} FWHM triangular bandpass & 3968.47\,{\AA} & 3969.59\,{\AA} \\
$\widetilde{K}_{\rm tri}$ & {\tt\string K\_mean\_tri} & Mean flux in $K$ band with 1.09\,{\AA} FWHM triangular bandpass & 3933.66\,{\AA} & 3934.78\,{\AA} \\
$\widetilde{H}_{\rm rec}$ & {\tt\string H\_mean\_rec} & Mean flux in $H$ band with 1\,{\AA} rectangular bandpass & 3968.47\,{\AA} & 3969.59\,{\AA} \\
$\widetilde{K}_{\rm rec}$ & {\tt\string K\_mean\_rec} & Mean flux in $K$ band with 1\,{\AA} rectangular bandpass & 3933.66\,{\AA} & 3934.78\,{\AA}
\enddata
\tablecomments{
`Column in Database' is used in Section \ref{sec:catalog}.
The wavelengths in air are taken from \citet{1978PASP...90..267V}.
The wavelengths in vacuum are calculated from air wavelengths using the formula given by \citet{1996ApOpt..35.1566C}.
}
\end{deluxetable}

\subsection{Evaluation of Chromospheric Activity Indexes}\label{sec:activity_indexes}

From the emission flux measures of Ca II H\&K lines obtained in Section \ref{sec:flux_measurements}, stellar chromospheric activity indexes can be evaluated.
The widely used chromospheric activity indicator, the classical MWO $S$ index (denoted by ${S_{\rm MWO}}$), was originally defined at MWO as \citep{1968ApJ...153..221W, 1978PASP...90..267V, 1991ApJS...76..383D, 1995ApJ...438..269B}
\begin{equation} \label{eq:S_MWO}
S_{\rm MWO}=\alpha \cdot \frac{N_{H}+N_{K}}{N_{R}+N_{V}},
\end{equation}
where $N_{H}$ and $N_{K}$ are the number of counts in the 1.09\,{\AA} FWHM triangular bandpasses of Ca II H and K lines,
$N_{R}$ and $N_{V}$ are the number of counts in the 20\,{\AA} $R$ and $V$ reference bands \citep{1978PASP...90..267V}, 
and $\alpha$ is a scaling factor for adjusting the HKP-2 measurements to be in the similar scale as HKP-1 results \citep{1991ApJS...76..383D}.

For the LAMOST LRS spectra, by referring to the definition of ${S_{\rm MWO}}$, we can introduce the LAMOST $S$ index (denoted by $S_L$) which is expressed as \citep{2011arXiv1107.5325L, 2016NatCo...711058K}
\begin{equation} \label{eq:S_L}
    S_L = \alpha_L \cdot \frac{8 \times 1.09\,\text{\AA}}{20\,\text{\AA}} \cdot \frac{\widetilde{H}_{\rm tri} + \widetilde{K}_{\rm tri}}{\widetilde{R}+\widetilde{V}},
\end{equation}
where $\alpha_L$ is the scaling factor for LAMOST, and $\widetilde{H}_\mathrm{tri}$, $\widetilde{K}_\mathrm{tri}$, $\widetilde{R}$, and $\widetilde{V}$ are the emission flux measures of Ca II H\&K lines defined in Section \ref{sec:flux_measurements} (see Table \ref{tab:CaIIHK-emission-flux-measures}).
The reason for multiplying 8 to the 1.09\,{\AA} FWHM is that the integration time spent on the $H$ and $K$ bands by the spectrometer used in MWO is eight times longer than that on the 20\,{\AA} wide $R$ and $V$ bands \citep{1991ApJS...76..383D, 2011arXiv1107.5325L}.
The value of $\alpha_L$ is adopted as 1.8 for LAMOST LRS data as suggested by \citet{2016NatCo...711058K}, which is determined based on the $S$-index distributions of solar-like stars.

\begin{figure}
    \epsscale{0.6}
	\plotone{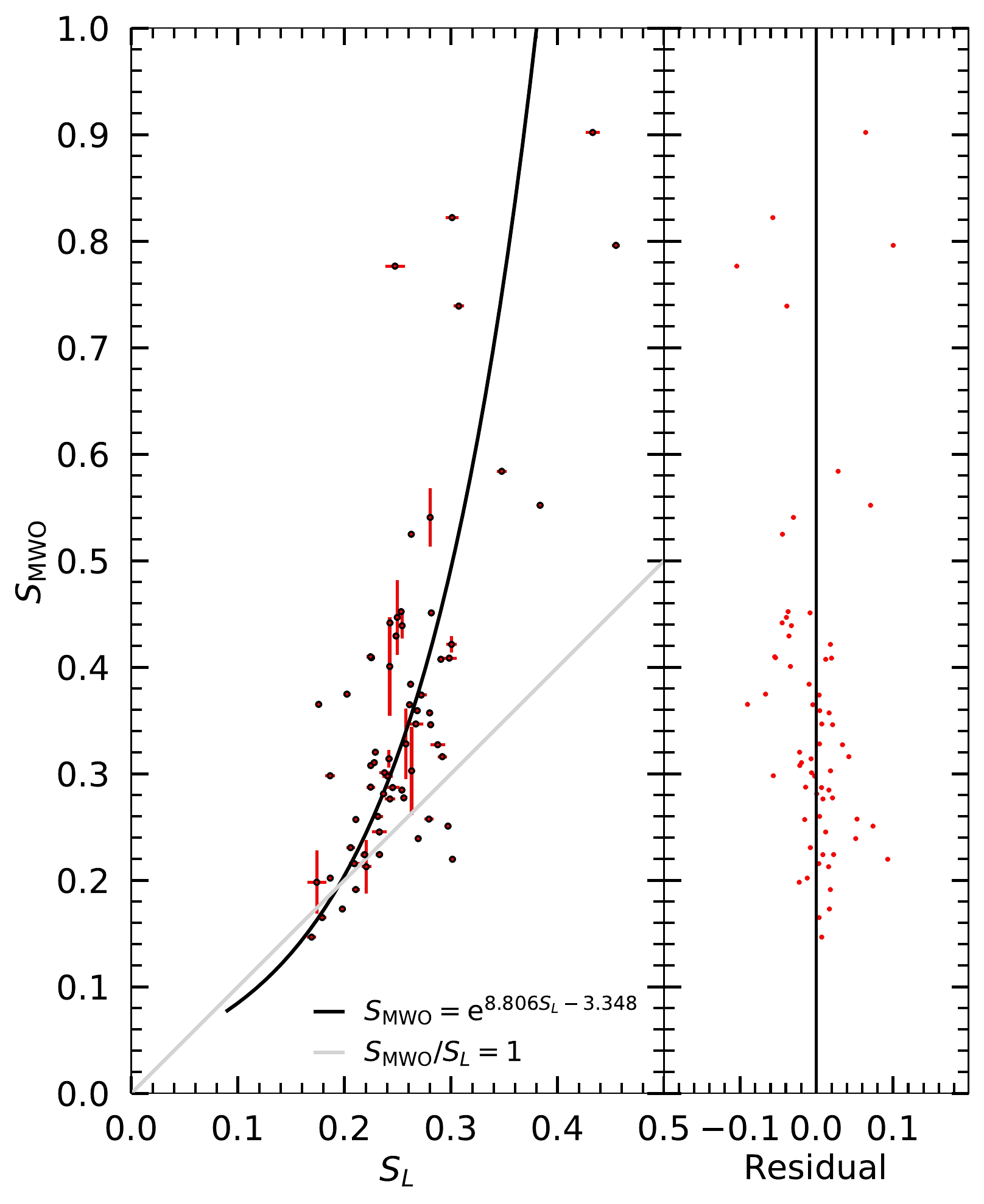}
	\caption{Scatter plot of $S_L$ vs. $S_{\rm MWO}$ for 65 common stars between the selected LRS spectra of solar-like stars in this work and the $S_{\rm MWO}$ catalog given by \citet{1991ApJS...76..383D}.
	Error bars are displayed for the data points with known uncertainty values. 
	The relationship between $S_L$ and $S_{\rm MWO}$ is fitted with an exponential function (black line; see Appendix \ref{sec:calibration_S_L_vs_S_MWO} for details of the common stars and fitting procedure). 
	The gray line indicates the ratio of $S_{\rm MWO}/S_L=1$. Note that the fitted line approaches $S_{\rm MWO}/S_L=1$ for smaller $S$-index values.
	\label{fig:S_MWO-S_L}
	}
\end{figure}

The relationship between the values of $S_L$ derived from LAMOST LRS spectra and the values of $S_{\rm MWO}$ measured by MWO can be calibrated based on the common stars of the two data sets.
In Figure \ref{fig:S_MWO-S_L}, we show the scatter plot of $S_L$ vs. $S_{\rm MWO}$ for 65 common stars between the selected LRS spectra of solar-like stars in this work and the $S_{\rm MWO}$ catalog given by \citet{1991ApJS...76..383D}.
As exhibited in Figure \ref{fig:S_MWO-S_L}, the relationship between $S_L$ and $S_{\rm MWO}$ can be fitted with an exponential function 
\begin{equation} \label{eq:S_L_vs_S_MWO}
    S_{\rm MWO} = \textrm{\large e}^{\,8.806\,S_L-3.348},
\end{equation}
which is displayed as a black line in Figure \ref{fig:S_MWO-S_L} (see Appendix \ref{sec:calibration_S_L_vs_S_MWO} for details of the common stars and fitting procedure).
The nonlinear relation is expected for larger $S$-index values since they are obtained by instruments with distinct spectral resolutions (see, e.g., \citealt{1978PASP...90..267V, 1996AJ....111..439H}, for more examples).
For smaller $S_{\rm MWO}$ values (less than 0.3), the fitted line approaches $S_{\rm MWO}/S_L=1$ as exhibited in Figure \ref{fig:S_MWO-S_L}, illustrating the suitability of the scaling factor value of $\alpha_L=1.8$ adopted for LAMOST LRS data.

The median of the relative deviations between the $S_L$ values of the common stars and the fitted line in Figure \ref{fig:S_MWO-S_L} is about 8.8\%. 
The large residuals could be due to the long-term activity variation in these stars (e.g., \citealt{1978ApJ...226..379W, 1990Natur.348..520B, 1995ApJ...438..269B, 2018ApJ...855...75R}), and LAMOST only captures a snapshot observation of these stars' activity at all $S$-index values.

If the activity index values are not needed to be in the similar scale as the measurements at MWO, the factors in Equation (\ref{eq:S_L}) are unnecessary.
Therefore, we can define the $S_{\rm tri}$ index
\begin{equation} \label{eq:S_tri}
    S_{\rm tri} = \frac{\widetilde{H}_{\rm tri} + \widetilde{K}_{\rm tri}}{\widetilde{R}+\widetilde{V}}
\end{equation}
based on the 1.09\,{\AA} FWHM triangular bandpass for Ca II H\&K lines. The $S_L$ index is connected with the $S_{\rm tri}$ index by equation
\begin{equation} \label{eq:S_L_vs_S_tri}
    S_L = 1.8 \times \frac{8 \times 1.09\,\text{\AA}}{20\,\text{\AA}} \times S_{\rm tri} \approx 0.785\,S_{\rm tri}.
\end{equation}

If it is not needed to keep the bandpass shape adopted by MWO, we can also use the 1\,{\AA} rectangular bandpasses to measure the line core emissions of Ca II H\&K lines (see Section \ref{sec:flux_measurements}) and define the $S_{\rm rec}$ index as
\begin{equation} \label{eq:S_rec}
    S_{\rm rec} = \frac{\widetilde{H}_{\rm rec} +\widetilde{K}_{\rm rec}}{\widetilde{R}+\widetilde{V}}.
\end{equation}

\citet{2011A&A...531A...8J} analyzed the influences of change in width of bandpasses at line cores of Ca II H\&K lines on the result of $S$ index.
Their result showed that the correlation between the derived $S$-index values and the original values of MWO decreases with increasing bandpass width (such as 2\,{\AA}, 3\,{\AA}, 4\,{\AA}, etc.) even for low-resolution spectral data.
Therefore, we adopt the current bandpass widths for defining the activity indexes $S_{\rm tri}$, $S_L$, and $S_{\rm rec}$ to keep best compatibility with the MWO measurements.

We calculated the values of $S_{\rm tri}$, $S_L$, and $S_{\rm rec}$ indexes by using Equations (\ref{eq:S_tri}), (\ref{eq:S_L_vs_S_tri}), and (\ref{eq:S_rec}) for all the selected LRS spectra of solar-like stars. 
The correlations between the values of the three indexes are illustrated and discussed in Section \ref{sec:tri_vs_rec}.

\subsection{Estimation of Uncertainties of Emission Flux Measures and Activity Indexes \label{sec:uncertainties}}

In Sections \ref{sec:flux_measurements} and \ref{sec:activity_indexes}, six emission flux measures of Ca II H\&K lines and three stellar chromospheric activity indexes are derived from the LRS spectra.
In this subsection, we estimate uncertainties of these activity parameters.
Three sources of uncertainty are taken into account: the uncertainty of spectral flux, the discretization in spectral data, and the uncertainty of radial velocity,
which can finally propagate into the composite uncertainty values of the derived activity parameters.

The FITS file of a LRS spectrum provides the value of inverse variance ($1/\delta_0^2$, 
where $\delta_0$ denotes the uncertainty of flux) to indicate photon noise for each data point in the spectrum.
As described in Sections \ref{sec:flux_measurements} and \ref{sec:activity_indexes}, the emission flux measures as well as the activity indexes are calculated based on the interpolated spectral data. 
The flux uncertainty value of an interpolated data point (denoted by $\delta_i$) can be derived from $\delta_0$ by equation
\begin{equation}\label{eq:delta_i-delta_0}
	\frac{\delta_i^2}{n_i} = \frac{\delta_0^2}{n_0},
\end{equation}
where $n_i$ is number density of the spectral data after interpolation and $n_0$ is number density in the original LRS spectrum. 
Then, for a given activity parameter $P$ (one of the six emission flux measures and three activity indexes), the uncertainty of $P$ caused by the uncertainty of spectral flux (denoted by ${\delta P}_{\rm flux}$) can be estimated from $\delta_i$ based on the definitions or formulas in Section \ref{sec:flux_measurements} and \ref{sec:activity_indexes} using the error propagation rules.

A spectrum is stored in discrete data points, and 
the flux values between data points have to be obtained via an interpolation algorithm as described in Section \ref{sec:flux_measurements}, 
which leads to uncertainties of the derived emission flux measures and activity indexes. 
To estimate the uncertainties of the activity parameters caused by the discretization in spectral data, we utilize two interpolation algorithm to obtain the spectral flux values between data points. 
One is the linear interpolation algorithm as used in Section \ref{sec:flux_measurements}, and another is the cubic interpolation algorithm.
For a given activity parameter $P$, we can get two parameter values, $P_{\rm linear}$ and $P_{\rm cubic}$, corresponding to the two interpolation algorithms, respectively. 
Then, the uncertainty of $P$ caused by the discretization in spectral data (denoted by $\delta P_{\rm discrete}$) can be estimated as the difference between $P_{\rm linear}$ and $P_{\rm cubic}$, i.e.,
\begin{equation} \label{eq:uncertainty_by_discretization}
	{\delta P}_{\rm discrete} = |P_{\rm cubic} - P_{\rm linear}|.
\end{equation}

To estimate the uncertainties of the emission flux measures and activity indexes caused by the uncertainty of radial velocity, we perform wavelength correction for a LRS spectrum (as described in Section \ref{sec:rv-correction}) using three deliberately set radial velocity values, $V_r - \delta V_r$, $V_r$, and $V_r + \delta V_r$, 
where $V_r$ is the formal radial velocity of the spectrum determined by LASP and $\delta V_r$ is the uncertainty of the radial velocity. 
For a given activity parameter $P$, we can get three parameter values, $P_-$, $P$, and $P_+$, corresponding to the three deliberately set radial velocity values, respectively. 
Then, the uncertainty of $P$ caused by the uncertainty of radial velocity (denoted by ${\delta P}_{V_r}$) can be estimated by the following formula:
\begin{equation} \label{eq:uncertainty_by_rv_err}
	{\delta P}_{V_r} = \frac{| P_- - P | + | P_+ - P |}{2}.
\end{equation}

The composite uncertainty of $P$ (denoted by $\delta P$) caused by all the three uncertainty sources can be calculated by formula
\begin{equation} \label{eq:composite_uncertainty}
	\delta P = \sqrt{({\delta P}_{\rm flux})^2+({\delta P}_{\rm discrete})^2+({\delta P}_{V_r})^2}.
\end{equation}

\begin{figure}
    \epsscale{0.6}
	\plotone{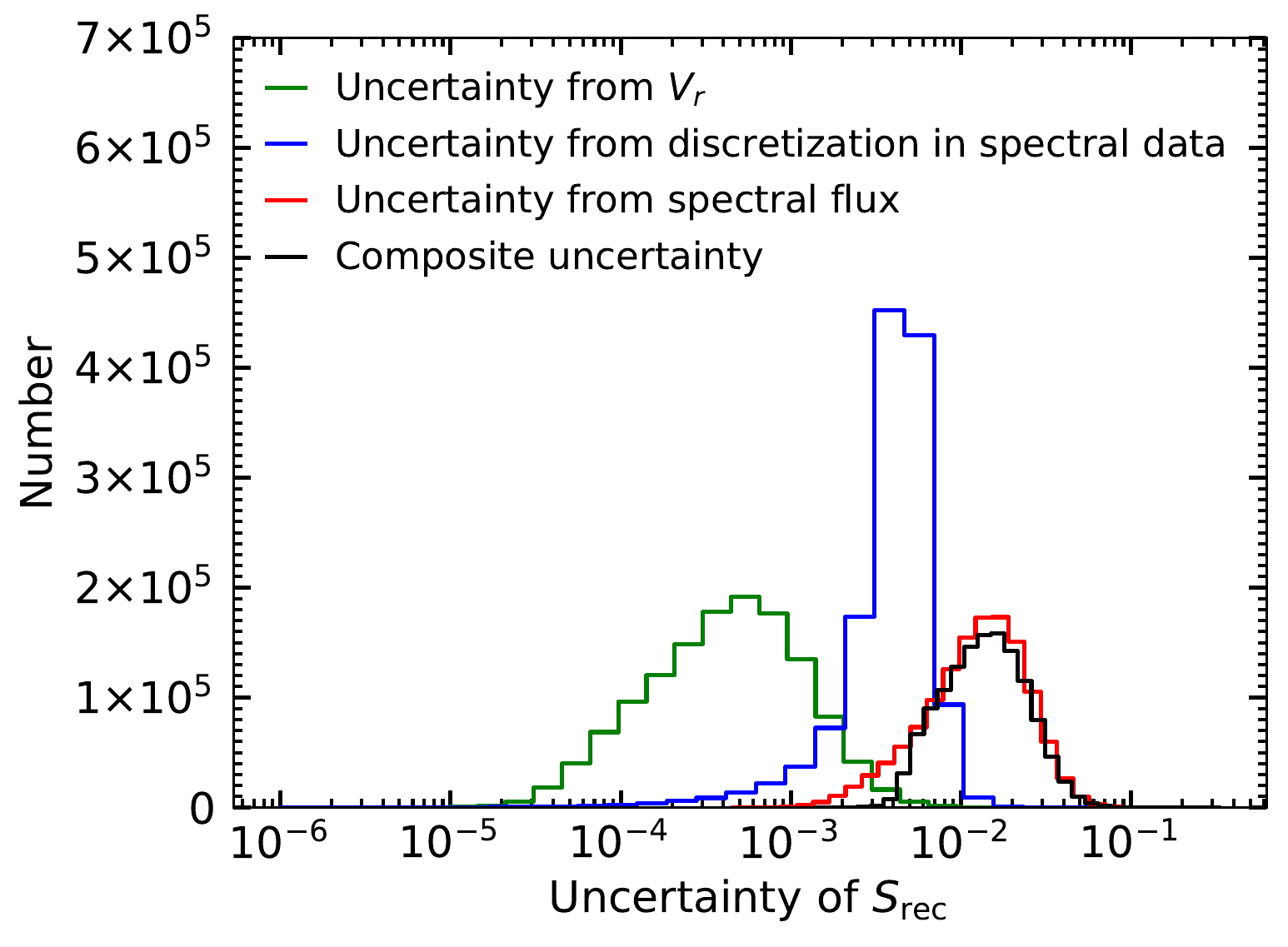}
	\caption{Distribution histograms of the uncertainties of the $S_{\rm rec}$ index originating from $V_r$ uncertainty (green line), discretization in spectral data (blue line), spectral flux uncertainty (red line), and all uncertainty sources (composite uncertainty; black line).
	\label{fig:S_rec_err_hist}
	}
\end{figure}

We calculate the uncertainty values (${\delta P}_{\rm flux}$, ${\delta P}_{V_r}$, and $\delta P$) of the six emission flux measures and three activity indexes for all the selected LRS spectra of solar-like stars.
In Figure \ref{fig:S_rec_err_hist}, we show the relative magnitudes among the uncertainties originating from different sources, by using the uncertainty of the activity index $S_{\rm rec}$ as an example.
As illustrated in Figure \ref{fig:S_rec_err_hist}, the uncertainty of $S_{\rm rec}$ originating from the uncertainty of spectral flux (about $10^{-2}$) is roughly an order of magnitude higher than that from the discretization in spectral data (about $10^{-3}$), which in turn is roughly an order of magnitude higher than that from the uncertainty of radial velocity (about $10^{-4}$).

All results of the obtained emission flux measures and activity indexes, and the estimated composite uncertainties of the activity parameters are integrated into the catalog of the stellar chromospheric activity database of solar-like stars (see Section \ref{sec:database}). 
In a few of LRS data, the inverse variance values are not available at some data points in the $H$, $K$, $R$, and $V$ bands of Ca II H\&K lines. For those spectra, the uncertainty values of the emission flux measures and activity indexes are filled with `{\tt\string -9999}' in the catalog of the database.

\section{Stellar Chromospheric Activity Database of Solar-like Stars}\label{sec:database}

In Section \ref{sec:lrs-spectra-selection}, we select 1,330,654 high-quality LRS spectra of solar-like stars from LAMOST DR7 v2.0. 
In Section \ref{sec:data-processing}, we derive six emission flux measures ($\widetilde{R}$, $\widetilde{V}$, $\widetilde{H}_{\rm tri}$, $\widetilde{K}_{\rm tri}$, $\widetilde{H}_{\rm rec}$, and $\widetilde{K}_{\rm rec}$) and three stellar chromospheric activity indexes ($S_{\rm tri}$, $S_L$, and $S_{\rm rec}$) of Ca II H\&K lines as well as their uncertainties for the selected LAMOST LRS spectra. 
These emission flux measures and activity indexes can be used to investigate the overall distribution of chromospheric activity of solar-like stars as well as the activity characteristics of individual spectra.
We also produce spectrum diagrams of Ca II H\&K lines for all the selected LRS spectra.
A stellar chromospheric activity database of solar-like stars are constructed based on the selected LAMOST LRS spectra, the derived emission flux measures and activity indexes, and the produced spectrum diagrams of Ca II H\&K lines.
The entity of the database is composed of a catalog of the spectral sample and activity parameters, and a library of the spectrum diagrams,
which are described in detail in the following subsections.
An online version of the database is available.\footnote{\url{https://doi.org/10.5281/zenodo.7067044}}
The original FITS data files of the associated LRS spectra can be queried and downloaded from the LAMOST website (see Section \ref{sec:lamost-dr}) through the {\tt\string obsid} or {\tt\string fitsname} information (see Table \ref{tab:catalog-columns}) included in the catalog the database.

\subsection{Catalog of Spectral Sample and Activity Parameters} \label{sec:catalog}
The catalog of the 1,330,654 high-quality LAMOST LRS spectra of solar-like stars as well as the derived emission flux measures and activity indexes is stored in a CSV format file (filename: {\tt\string CaIIHK\_Sindex\_LAMOST\_DR7\_LRS.csv}). 
Each row of the catalog corresponds to a LRS spectrum. 
All the columns in the catalog are tabulated in Table \ref{tab:catalog-columns} with brief descriptions. 
As shown in Table \ref{tab:catalog-columns}, the six emission flux measures (columns: {\tt\string R\_mean}, {\tt\string V\_mean}, {\tt\string H\_mean\_tri}, {\tt\string K\_mean\_tri}, {\tt\string H\_mean\_rec}, and {\tt\string K\_mean\_rec}), 
the three activity indexes (columns: {\tt\string S\_tri}, {\tt\string S\_L}, and {\tt\string S\_rec}), and their uncertainties derived in Section \ref{sec:data-processing} are included in the catalog (18 columns in total).
The figure file name of Ca II H\&K spectrum diagram (see Section \ref{sec:spectrum-diagram-library}) is also included in the catalog (column: {\tt\string figname}).
The aforementioned 19 columns provided by this work are labeled with a `\textasteriskcentered' symbol in Table \ref{tab:catalog-columns}. 
Other columns in Table \ref{tab:catalog-columns} are taken from the data release of LAMOST; 
those columns are used in this work and hence are kept in the catalog for reference.

\startlongtable
\begin{deluxetable}{llllll}
	\tablecaption{Columns in the catalog of the database.\label{tab:catalog-columns}}
	\tablehead{
		& \colhead{Column} && \colhead{Unit} && \colhead{Description} 
	}
	\startdata
	& {\tt\string obsid} &&  && Observation identifier of LAMOST spectrum\\
	& {\tt\string fitsname} && && FITS file name of LAMOST spectrum\\
	& {\tt\string snrg} && && SNR in $g$ band ($\mathrm{SNR}_g$) \\
	& {\tt\string snrr} && && SNR in $r$ band ($\mathrm{SNR}_r$)\\
	& {\tt\string teff} && K && Effective temperature ($T_\mathrm{eff}$)\\
	& {\tt\string teff\_err} && K && Uncertainty of $T_\mathrm{eff}$\\
	& {\tt\string logg} && dex && Surface gravity ($\log\,g$) \\
	& {\tt\string logg\_err} && dex && Uncertainty of $\log\,g$ \\
	& {\tt\string feh} && dex && Metallicity ([Fe/H]) \\
	& {\tt\string feh\_err} && dex && Uncertainty of [Fe/H] \\
	& {\tt\string rv} && km/s && Radial velocity ($V_r$)\\
	& {\tt\string rv\_err} && km/s && Uncertainty of $V_r$\\
	& {\tt\string ra\_obs} && degree && Right ascension (RA) of fiber pointing \\
	& {\tt\string dec\_obs} && degree && Declination (DEC) of fiber pointing \\
	& {\tt\string gaia\_source\_id} && && Source identifier in Gaia DR2 catalog\\
	& {\tt\string gaia\_g\_mean\_mag} && && $G$ magnitude in Gaia DR2 catalog\\
	& {\tt\string figname}* && && Figure file name of Ca II H\&K spectrum diagram \\
	& {\tt\string R\_mean}* && && Mean flux in $R$ band ($\widetilde{R}$) \\
	& {\tt\string R\_mean\_err}* && && Uncertainty of $\widetilde{R}$\\ 
	& {\tt\string V\_mean}* && && Mean flux in $V$ band ($\widetilde{V}$) \\
	& {\tt\string V\_mean\_err}* && && Uncertainty of $\widetilde{V}$ \\
    & {\tt\string H\_mean\_tri}* && && Mean flux in $H$ band with 1.09\,{\AA} FWHM triangular bandpass ($\widetilde{H}_{\rm tri}$) \\ 
	& {\tt\string H\_mean\_tri\_err}* && && Uncertainty of $\widetilde{H}_{\rm tri}$ \\ 
	& {\tt\string K\_mean\_tri}* && && Mean flux in $K$ band with 1.09\,{\AA} FWHM triangular bandpass ($\widetilde{K}_{\rm tri}$) \\
	& {\tt\string K\_mean\_tri\_err}* && && Uncertainty of $\widetilde{K}_{\rm tri}$ \\
	& {\tt\string S\_tri}* && && $S_{\rm tri}$ index using 1.09\,{\AA} FWHM triangular bandpasses of $H$ and $K$ bands \\
	& {\tt\string S\_tri\_err}* && && Uncertainty of $S_{\rm tri}$ \\
	& {\tt\string S\_L}* && && LAMOST $S$ index ($S_L$) \\
	& {\tt\string S\_L\_err}* && && Uncertainty of $S_L$\\
	& {\tt\string H\_mean\_rec}* && && Mean flux in $H$ band with 1\,{\AA} rectangular bandpass ($\widetilde{H}_{\rm rec}$) \\ 
	& {\tt\string H\_mean\_rec\_err}* && && Uncertainty of $\widetilde{H}_{\rm rec}$ \\ 
	& {\tt\string K\_mean\_rec}* && && Mean flux in $K$ band with 1\,{\AA} rectangular bandpass ($\widetilde{K}_{\rm rec}$) \\
	& {\tt\string K\_mean\_rec\_err}* && && Uncertainty of $\widetilde{K}_{\rm rec}$ \\
	& {\tt\string S\_rec}* && && $S_{\rm rec}$ index using 1\,{\AA} rectangular bandpasses of $H$ and $K$ bands \\
	& {\tt\string S\_rec\_err}* && && Uncertainty of $S_{\rm rec}$ \\
	\enddata
	\tablecomments{Columns labeled with a `\textasteriskcentered' symbol are provided by this work. Other columns are used in this work and are taken from the data release of LAMOST.}
\end{deluxetable}

\subsection{Library of Spectrum Diagrams} \label{sec:spectrum-diagram-library}
\begin{figure}
    \epsscale{1.09}
	\plotone{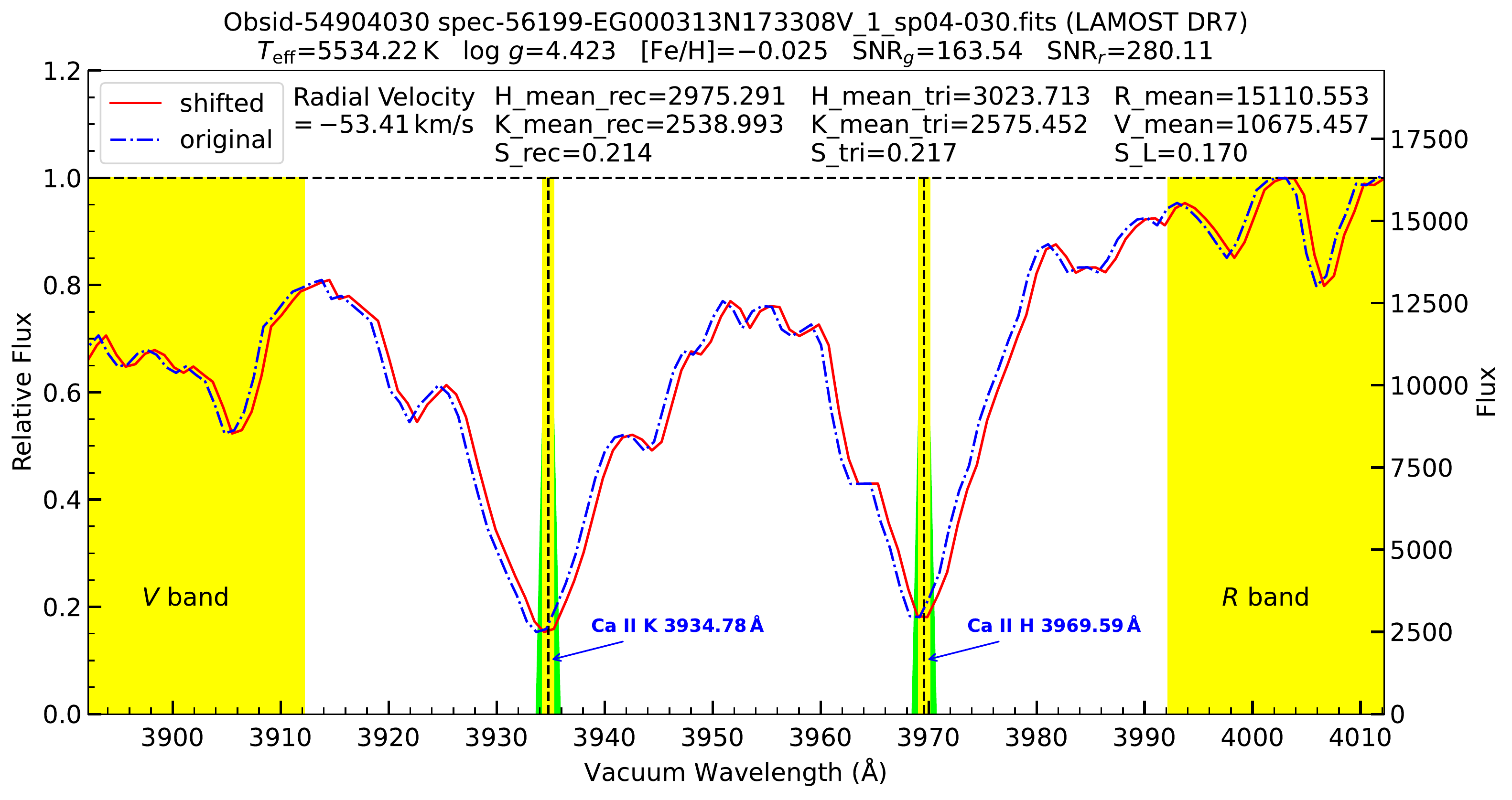}
	\caption{An example of the spectrum diagrams of Ca II H\&K lines in the database using a LAMOST LRS spectrum with $\text{obsid}=54904030$. 
	The wavelength range of the diagram is 3892.17--4012.20\,{\AA}.
	The original flux of the spectrum released by LAMOST is shown through the right-hand vertical axis. 
	The left-hand vertical axis is the relative flux normalized by the maximum value of the original flux in the plot.
	The blue dash-dot line is the original LAMOST spectrum, and the red solid line is the wavelength-shifted spectrum after radial velocity correction.
	The radial velocity value of the spectrum is given in the upper left area of the diagram.
	The two yellow rectangular regions on the two sides of the diagram indicate the 20\,{\AA} wide $R$ and $V$ pseudo-continuum bands. 
	The two green triangular regions centered at 3969.59\,{\AA} and 3934.78\,{\AA} indicate the 1.09\,{\AA} FWHM triangular bandpasses for Ca II H line and K line, respectively. 
	Within the triangular regions are the {1\,\AA} rectangular bandpasses which are shown in yellow.
	The stellar chromospheric activity parameters derived in this work are displayed in the area just above the spectrum plot.
	The obsid, FITS file name, data release number, and several stellar and spectroscopic parameters of the LAMOST spectrum are given in the title area of the diagram.
	\label{fig:spectrum_diagram_example}}
\end{figure}

To intuitively illustrate spectral profiles, we produce spectrum diagram of Ca II H\&K lines of LAMOST LRS spectra.
Figure \ref{fig:spectrum_diagram_example} shows an example of the spectrum diagrams using a LRS spectrum with $\text{obsid}=54904030$.
In the spectrum diagram, the original LRS spectrum is displayed with a blue dash-dot line, and the wavelength-shifted spectrum after radial velocity correction (see Section \ref{sec:rv-correction}) is displayed with a red solid line.
The right-hand vertical axis shows the original flux of the spectrum released by LAMOST. 
This original flux is normalized by its maximum value in the wavelength range of the diagram (3892.17--4012.20\,{\AA}) to yield relative flux. The relative flux after normalization is displayed in the left-hand vertical axis.

The 20\,{\AA} wide $R$ and $V$ pseudo-continuum bands are illustrated by two yellow rectangular regions on the right side and left side of the spectrum diagram, respectively.
The 1.09\,{\AA} FWHM triangular $H$ and $K$ bands are indicated by two green triangular regions centered at 3969.59\,{\AA} and 3934.78\,{\AA} (wavelengths of the Ca II H and K line cores in vacuum; see Table \ref{tab:CaIIHK-emission-flux-measures}), respectively.
Within the triangular regions are the {1\,\AA} rectangular $H$ and $K$ bands which are shown in yellow.
In the title area of the spectrum diagram, the obsid, FITS file name, and data release number of the LAMOST spectrum are given in the first line, and several stellar and spectroscopic parameters used in this work ($T_{\rm eff}$, $\log\,g$, [Fe/H], $\mathrm{SNR}_g$, and $\mathrm{SNR}_r$) are displayed in the second line.
The stellar chromospheric activity parameters derived in this work ({\tt\string H\_mean\_rec}, {\tt\string H\_mean\_tri}, {\tt\string R\_mean}, {\tt\string K\_mean\_rec}, {\tt\string K\_mean\_tri}, {\tt\string V\_mean}, {\tt\string S\_rec}, {\tt\string S\_tri}, and {\tt\string S\_L}) are shown in the area just above the spectrum plot.
The radial velocity value can be seen in the upper left area of the diagram.

We produced spectrum diagrams for all the 1,330,654 LAMOST LRS spectra of solar-like stars employed by the database. 
These spectrum diagrams are saved as JPG format figures. 
All the figures constitute the library of spectrum diagrams of the database. 
The filenames of the figures follow the convention of `{\tt\string CaIIHK\_obsid-}\textlangle{\it 9-digit-obsid}\textrangle{\tt\string _}\textlangle{\it primary name of FITS file}\textrangle{\tt\string .jpg}' and are included in the catalog of the database (`{\tt\string figname}' in Table \ref{tab:catalog-columns}).
To facilitate retrieval and utilization of the spectrum diagrams, we rewrite LAMOST obsid as a 9-digit number.
If the number of digits of an original obsid is less than 9, `0' is padded on the left. (For example, `103019' is rewritten as `000103019'). 
Diagrams with the same first three digits of the 9-digit-obsid are gathered into a folder named with the first three digits (such as `{\tt\string 000/}', `{\tt\string 001/}', etc.). 
The figure file of a certain spectrum diagram can be found through the path `\textlangle{\it first three digits of 9-digit-obsid}\textrangle{\tt\string /}\textlangle{\it figure file name of spectrum diagram}\textrangle'. 
For example, the figure file of the spectrum diagram displayed in Figure \ref{fig:spectrum_diagram_example} is located at `{\tt\string 054/CaIIHK\_obsid-054904030\_spec-56199-EG000313N173308V\_1\_sp04-030.jpg}'.
There are 585 folders in total. 
These folders of spectrum diagrams are archived into multiple compressed ZIP packages, with each ZIP package containing no more than 50 folders. 
The filenames of the ZIP files follow the convention such as `{\tt\string spectrum\_diagrams\_000-049.zip}' to show the range of the folders contained in the ZIP files.
These ZIP packages of spectrum diagrams are available through the online version of the database.

\section{Results and Discussion} \label{sec:results}

\subsection{Triangular Bandpass versus Rectangular Bandpass of Ca II H and K Lines} \label{sec:tri_vs_rec}

\begin{figure}
    \epsscale{1.14}
	\plotone{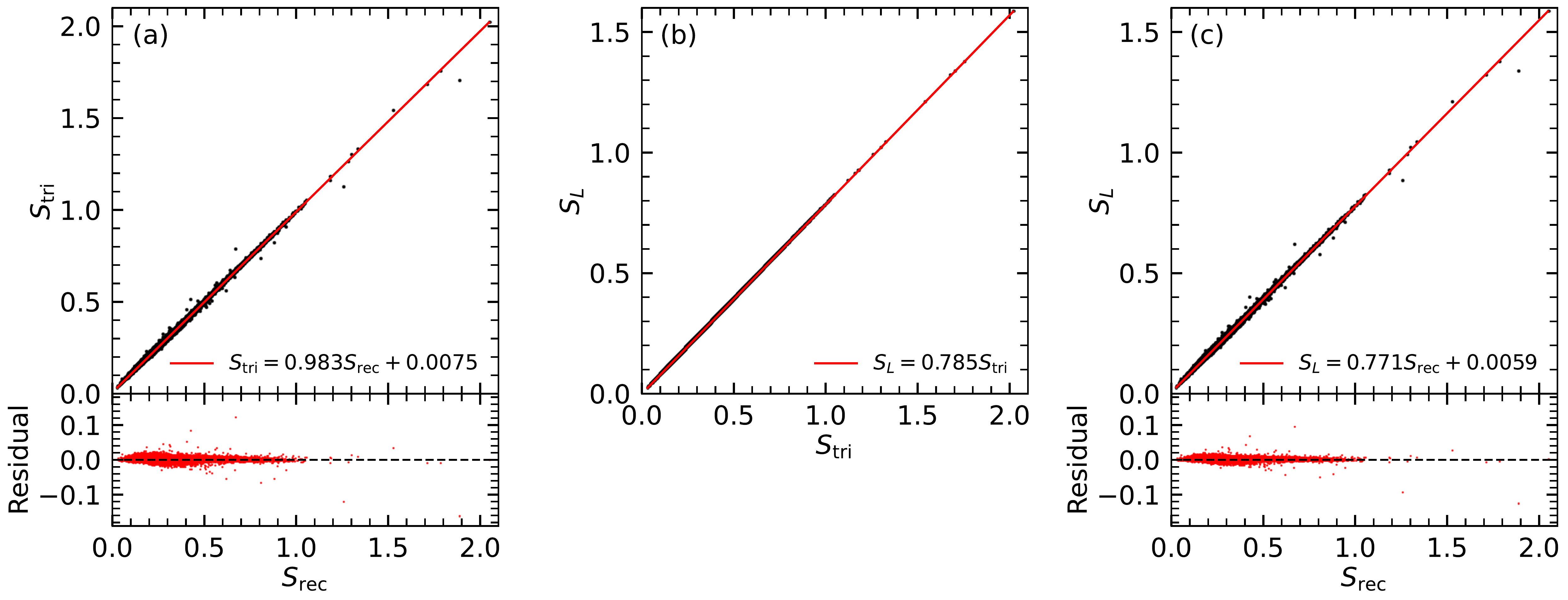}
	\caption{Scatter plots of (a) $S_{\rm rec}$ vs. $S_{\rm tri}$, (b) $S_{\rm tri}$ vs. $S_L$, and (c) $S_{\rm rec}$ vs. $S_L$ based on the values of the activity indexes in the database. The red lines are linear fittings to the correlations between the activity indexes. The formulas for the fitting results are given in the plots. The residuals of the fittings are also given for panels (a) and (c). There are no residuals in panel (b) since the relation between $S_{\rm tri}$ and $S_L$ is directly defined by Equation (\ref{eq:S_L_vs_S_tri}).
	}
	\label{fig:S_L-Srec-Stri}
\end{figure}

In Section \ref{sec:activity_indexes}, the $S_{\rm tri}$ and $S_{\rm rec}$ indexes are introduced based on the triangular bandpass and the rectangular bandpass at Ca II H\&K line cores, respectively. 
The $S_L$ index is further introduced by multiplying $S_{\rm tri}$ by a scaling factor.
Figure \ref{fig:S_L-Srec-Stri} depicts the correlations of $S_{\rm rec}$ vs. $S_{\rm tri}$, $S_{\rm tri}$ vs. $S_L$, and $S_{\rm rec}$ vs. $S_L$ based on the derived values of the activity indexes in the database.

As shown in Figure \ref{fig:S_L-Srec-Stri}, there is a good consistency between the values of $S_{\rm rec}$, $S_{\rm tri}$, and  $S_L$ for the LRS spectra.
A linear fitting for $S_{\rm rec}$ vs. $S_{\rm tri}$ gives
\begin{equation} \label{eq:S_tri_vs_S_rec}
    S_{\rm tri} = 0.983\,S_{\rm rec} + 0.0075
\end{equation}
(see Figure \ref{fig:S_L-Srec-Stri}a), which means that for larger values ($>0.456$) of activity indexes, $S_{\rm rec}$ is generally slightly greater than $S_{\rm tri}$.
The relation between $S_{\rm tri}$ and $S_L$ has been given by Equation (\ref{eq:S_L_vs_S_tri}) (also see Figure \ref{fig:S_L-Srec-Stri}b).
A linear fitting for $S_{\rm rec}$ vs. $S_L$ gives 
\begin{equation} \label{eq:S_L_vs_S_rec}
    S_L = 0.771\,S_{\rm rec} + 0.0059
\end{equation}
(see Figure \ref{fig:S_L-Srec-Stri}c).
Note that Equation (\ref{eq:S_L_vs_S_rec}) can be deduced from Equations (\ref{eq:S_L_vs_S_tri}) and (\ref{eq:S_tri_vs_S_rec}).

In comparison to the $S_{\rm tri}$ and $S_L$ indexes defined based on the triangular bandpass at line cores of Ca II H\&K lines, the $S_{\rm rec}$ index defined based on the rectangular bandpass has a more straightforward physical meaning and is also a reliable and suitable choice for stellar activity studies with the LRS spectra.
The mean $S_{\rm MWO}$ index value of the Sun is about 0.169 as determined by \citet{2017ApJ...835...25E} based on the MWO HKP-2 measurements. 
By substituting this value to Equation (\ref{eq:S_L_vs_S_MWO}) and then using Equation (\ref{eq:S_L_vs_S_rec}), we can get the mean $S_{\rm rec}$ value of the Sun (denoted by $\langle S_{\rm rec,\,\odot}\rangle$) is about 0.223.

\subsection{Uncertainty of Activity Index versus SNR of Spectra} \label{sec:uncertainty_vs_snr}
The uncertainty of the activity indexes is related to the SNR of spectra. 
A larger SNR usually corresponds to a smaller uncertainty of activity index. 
We utilize all the derived uncertainty values of $S_{\rm rec}$ in Section \ref{sec:uncertainties} and the ${\rm SNR}_g$ values of the LAMOST LRS spectra to quantitatively analyze this relation.
(The analysis can also be performed for $S_{\rm tri}$ and $S_L$, and the results are similar.)

\begin{figure}
    \epsscale{0.74}
	\plotone{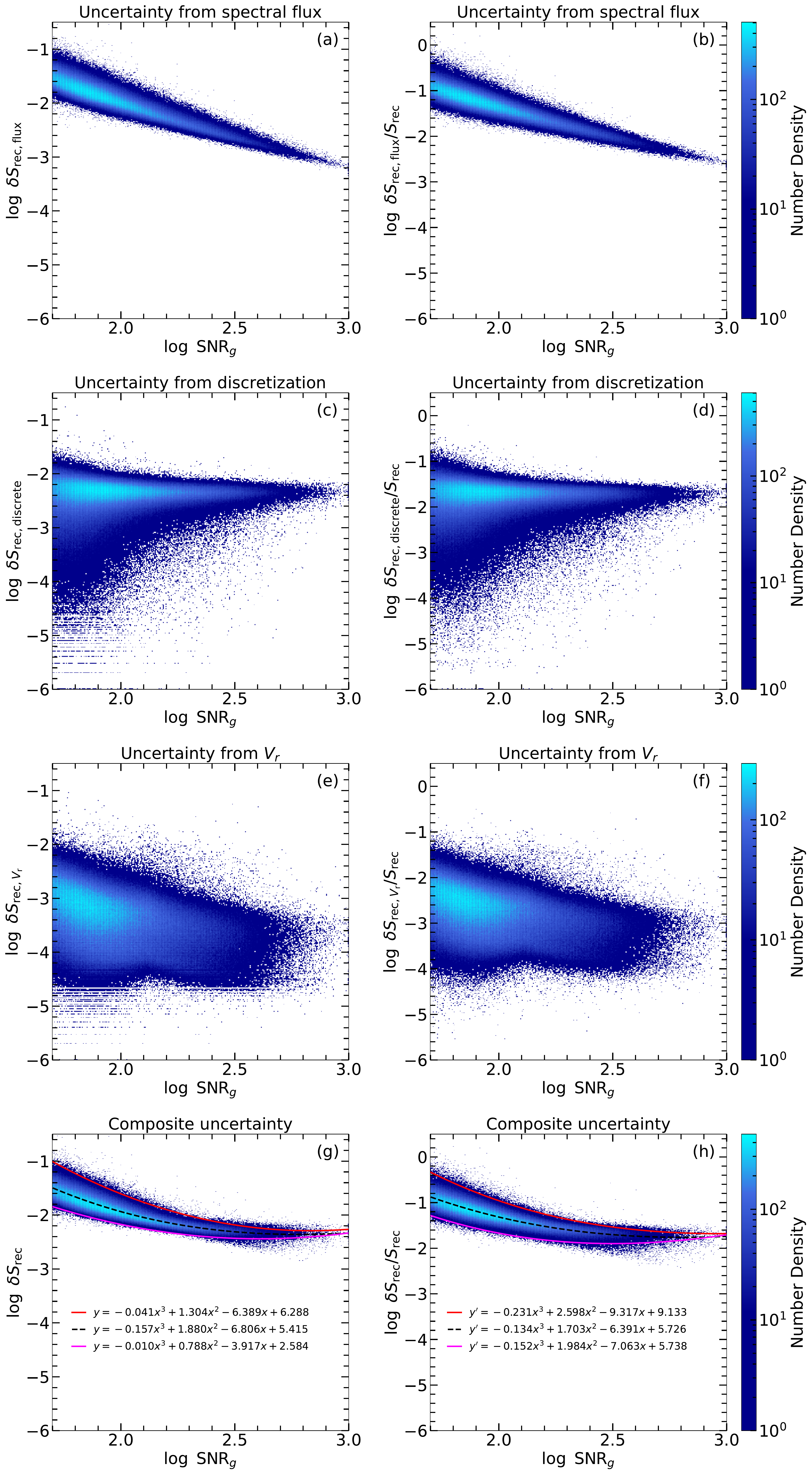}
	\caption{Distributions of $\log\,{\rm SNR}_g$ vs.
	(a) $\log\,\delta{S_{\rm rec,\,flux}}$, 
	(b) $\log\,\delta{S_{\rm rec,\,flux}}/{S_{\rm rec}}$, 
	(c) $\log\,\delta{S_{\rm rec,\,discrete}}$, 
	(d) $\log\,\delta{S_{\rm rec,\,discrete}}/{S_{\rm rec}}$, 
	(e) $\log\,\delta{S_{{\rm rec},\,V_r}}$, 
	(f) $\log\,\delta{S_{{\rm rec},\,V_r}}/{S_{\rm rec}}$, 
	(g) $\log\,\delta{S_{\rm rec}}$, and 
	(h) $\log\,\delta{S_{\rm rec}}/{S_{\rm rec}}$.
	The color scale represents number density.
	In panels g and h, the cubic polynomial fitting lines for upper envelope, mean value, and lower envelope of the distributions are plotted with red, black dashed, and violet lines, respectively.
	The formulas for the fitted lines are also displayed, with $x$ representing $\log\,{\rm SNR}_g$, $y$ representing $\log\,\delta{S_{\rm rec}}$, and $y'$ representing $\log\,\delta{S_{\rm rec}}/{S_{\rm rec}}$.
	}
	\label{fig:SNRg-S_rec_err}
\end{figure}

In Section \ref{sec:uncertainties}, we have obtained the uncertainties of $S_{\rm rec}$ originating from the uncertainty of spectral flux, the discretization in spectral data, and the uncertainty of radial velocity (denoted by $\delta{S_{\rm rec,\,flux}}$, $\delta{S_{\rm rec,\,discrete}}$, and $\delta{S_{{\rm rec},\,V_r}}$, respectively), as well as the composite uncertainty of $S_{\rm rec}$ caused by all the three uncertainty sources (denoted by $\delta{S_{\rm rec}}$).
The scatter plots illustrating the distributions of $\log\,{\rm SNR}_g$ vs. $\log\,\delta{S_{\rm rec,\,flux}}$, $\log\,{\rm SNR}_g$ vs. $\log\,\delta{S_{\rm rec,\,discrete}}$, $\log\,{\rm SNR}_g$ vs. $\log\,\delta{S_{{\rm rec},\,V_r}}$, and $\log\,{\rm SNR}_g$ vs. $\log\,\delta{S_{\rm rec}}$ are displayed in the left panels of Figure \ref{fig:SNRg-S_rec_err}. 
We also calculate the corresponding relative uncertainties of $S_{\rm rec}$ (i.e., ${\delta{S_{\rm rec,\,flux}}}/{S_{\rm rec}}$, ${\delta{S_{\rm rec,\,discrete }}}/{S_{\rm rec}}$, ${\delta{S_{{\rm rec},\,V_r}}}/{S_{\rm rec}}$, and ${\delta{S_{\rm rec}}}/{S_{\rm rec}}$), and the distributions of $\log\,{\rm SNR}_g$ vs. $\log\,{\delta{S_{\rm rec,\,flux}}}/{S_{\rm rec}}$, $\log\,{\rm SNR}_g$ vs. $\log\,{\delta{S_{\rm rec,\,discrete}}}/{S_{\rm rec}}$, $\log\,{\rm SNR}_g$ vs. $\log\,{\delta{S_{{\rm rec},\,V_r}}}/{S_{\rm rec}}$, and $\log\,{\rm SNR}_g$ vs. $\log\,{\delta{S_{\rm rec}}}/{S_{\rm rec}}$ are displayed in the right panels of Figure \ref{fig:SNRg-S_rec_err}.

It can be seen from Figure \ref{fig:SNRg-S_rec_err} that a power law relation is roughly satisfied between ${\rm SNR}_g$ and $\delta{S_{\rm rec,\,flux}}$ as well as between ${\rm SNR}_g$ and ${\delta{S_{\rm rec,\,flux}}}/{S_{\rm rec}}$ (Figures \ref{fig:SNRg-S_rec_err}a and b), 
however, the values of $\delta{S_{\rm rec}}$ and ${\delta{S_{\rm rec}}}/{S_{\rm rec}}$ are distributed over a relatively wide strip area.
The order of magnitude of $\delta{S_{\rm rec,\,discrete}}$ (Figures \ref{fig:SNRg-S_rec_err}c and d) is generally smaller than $\delta{S_{\rm rec,\,flux}}$, and the order of magnitude of $\delta{S_{{\rm rec},\,V_r}}$ (Figures \ref{fig:SNRg-S_rec_err}e and f) in turn is generally smaller than $\delta{S_{\rm rec,\,discrete}}$, which has been demonstrated in Figure \ref{fig:S_rec_err_hist}.

Figures \ref{fig:SNRg-S_rec_err}g and h show that the composite uncertainty $\delta{S_{\rm rec}}$ is mainly affected by $\delta{S_{\rm rec,\,flux}}$ for smaller ${\rm SNR}_g$ values and by $\delta{S_{\rm rec,\,discrete}}$ for larger ${\rm SNR}_g$ values.
We performed cubic polynomial fitting for the upper envelope, mean value, and lower envelope of the distributions of $\log\,\delta{S_{\rm rec}}$ and $\log\,{\delta{S_{\rm rec}}}/{S_{\rm rec}}$ in Figures \ref{fig:SNRg-S_rec_err}g and h.
We divide the range of $\log\,{\rm SNR}_g$ (from $\log 50$=1.7 to $\log 1000$=3.0) into 300 equal-width bins and use the upper envelope, mean, and lower envelope values of $\log\,\delta{S_{\rm rec}}$ and $\log\,{\delta{S_{\rm rec}}}/{S_{\rm rec}}$ in each bin for fitting. 
The upper envelope and lower envelope threshold is defined as the number density value in Figures \ref{fig:SNRg-S_rec_err}g and h no less than 5 in each bin. 
If all the number density values in a bin are less than 5, the bin does not participate in the fitting.
The fitting results are illustrated in Figures\ref{fig:SNRg-S_rec_err}g and h.

The formulas for the fitted upper envelope, mean value, and lower envelope of the $\log\,\delta{S_{\rm rec}}$ distribution are given in Equations (\ref{eq:s_err-snrg_1}), (\ref{eq:s_err-snrg_2}), and (\ref{eq:s_err-snrg_3}), respectively, in which $x$ represents $\log\,{\rm SNR}_g$ and $y$ represents $\log\,\delta{S_{\rm rec}}$:

\begin{equation} \label{eq:s_err-snrg_1}
	y = -0.041 x^3 + 1.304 x^2 - 6.389 x + 6.288,
\end{equation}
\begin{equation} \label{eq:s_err-snrg_2}
	y = -0.157 x^3 + 1.880 x^2 -6.806 x + 5.415,
\end{equation}
\begin{equation} \label{eq:s_err-snrg_3}
	y = -0.010 x^3 + 0.788 x^2 - 3.917 x + 2.584.
\end{equation}

The formulas for the fitted upper envelope, mean value, and lower envelope of the $\log\,{\delta{S_{\rm rec}}}/{S_{\rm rec}}$ distribution are given in Equations (\ref{eq:s_err-snrg_4}), (\ref{eq:s_err-snrg_5}), and (\ref{eq:s_err-snrg_6}), respectively, in which $x$ represents $\log\,{\rm SNR}_g$ and $y'$ represents $\log\,\delta{S_{\rm rec}}/{S_{\rm rec}}$:

\begin{equation}\label{eq:s_err-snrg_4}
	y' = - 0.231 x^3 + 2.598 x^2 - 9.317 x + 9.133,
\end{equation}
\begin{equation}\label{eq:s_err-snrg_5}
	y' = -0.134 x^3 + 1.703 x^2 - 6.391 x + 5.726,
\end{equation}
\begin{equation}\label{eq:s_err-snrg_6}
	y' = - 0.152 x^3 + 1.984 x^2 -7.063 x + 5.738.
\end{equation}

Equations (\ref{eq:s_err-snrg_1})--(\ref{eq:s_err-snrg_6}) can be used to make a preliminary estimation of the values of $\delta{S_{\rm rec}}$ and ${\delta{S_{\rm rec}}}/{S_{\rm rec}}$ from the value of $\mathrm{SNR}_g$.
For example, by using Equation (\ref{eq:s_err-snrg_5}) (illustrated by the black dashed line in Figure \ref{fig:SNRg-S_rec_err}h), it can be deduced that the mean value of $\log\,{\delta{S_{\rm rec}}}/{S_{\rm rec}}$ for ${\rm SNR}_g = 50$ is about $-0.87$ and the corresponding ${\delta{S_{\rm rec}}}/{S_{\rm rec}}$ value is about $10^{-0.87} \approx 0.13$.
By using Equation (\ref{eq:s_err-snrg_1}) (illustrated by the red line in Figure \ref{fig:SNRg-S_rec_err}g), it can be deduced that the upper envelope value of $\log\,\delta{S_{\rm rec}}$ for ${\rm SNR}_g = 50$ is about $-1.0$, 
which means the whole upper envelope line of $\log\,\delta{S_{\rm rec}}$ is well below $-1.0$ and the corresponding $\delta{S_{\rm rec}}$ value is below $10^{-1.0}=0.1$. 
Considering the distribution range of the $S_{\rm rec}$ values (about $10^0=1.0$; see Figures \ref{fig:S_rec_hist}), it is appropriate to set the upper limit of the uncertainty values of $S_{\rm rec}$ to about $10^{-1}=0.1$. This requirement leads to the ${\rm SNR}_g$ condition (${\rm SNR}_g \ge 50.00$) for selecting LRS spectra in 
Section \ref{sec:lrs-spectra-selection}.

\subsection{Overall Distribution of Chromospheric Activity of Solar-like Stars} \label{sec:sindex_dist}

We use the full data set of the derived $S_{\rm rec}$ values in the database to show the overall distribution of chromospheric activity of solar-like stars.
The distributions of $S_{\rm tri}$ and $S_L$ are similar since they have approximate linear relations with $S_{\rm rec}$ (see Figures \ref{fig:S_L-Srec-Stri}a and c).

\begin{figure}
    \epsscale{0.80}
	\plotone{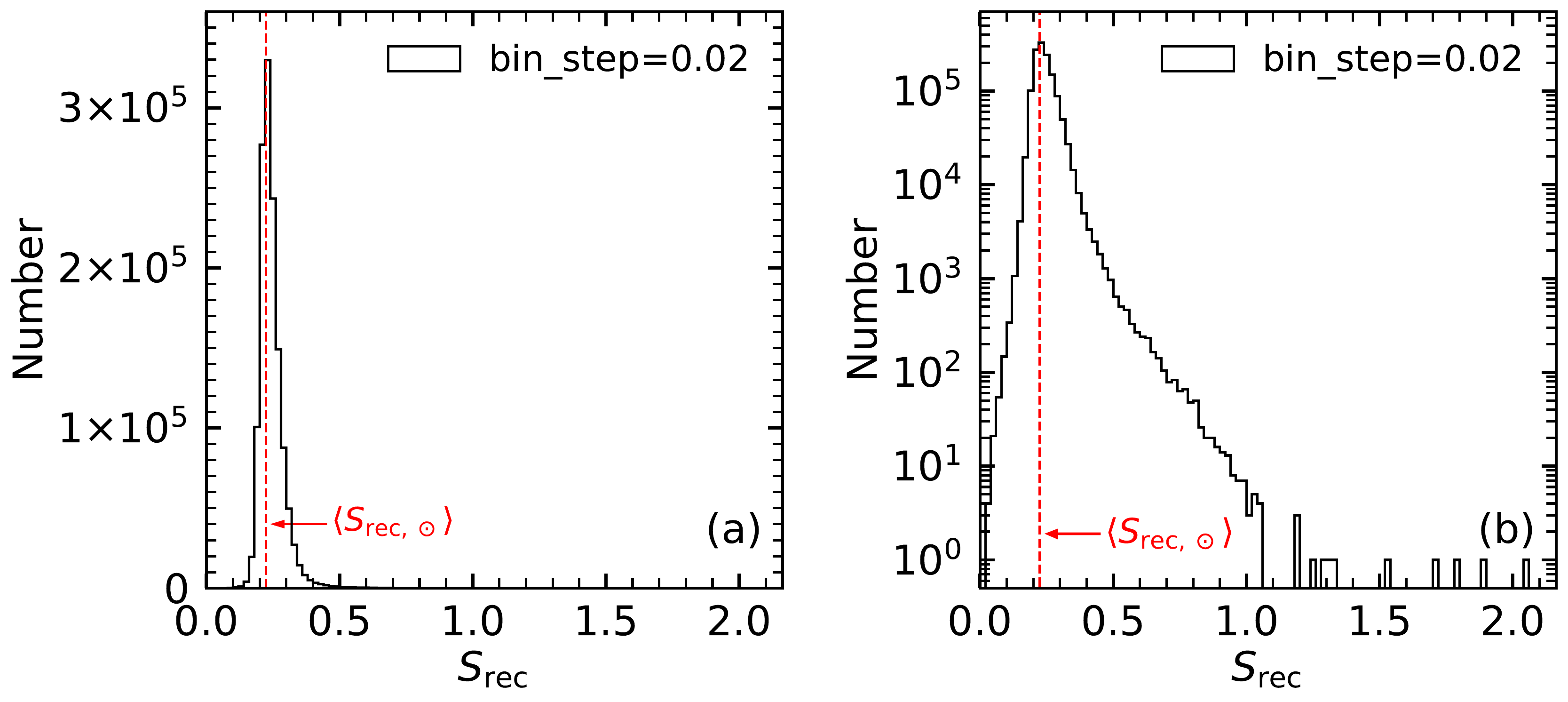}
	\caption{Histograms of the $S_{\rm rec}$ values of solar-like stars in the database with vertical axis (a) in linear scale and (b) in logarithmic scale. The mean $S_{\rm rec}$ value of the Sun ($\langle S_{\rm rec,\,\odot} \rangle = 0.223$) is indicated by a vertical dashed line in the plots.
	\label{fig:S_rec_hist}
	}
\end{figure}

The histogram of all the $S_{\rm rec}$ values in the database is displayed in Figure \ref{fig:S_rec_hist}. Figure \ref{fig:S_rec_hist}a uses a linear scale for vertical axis and Figure \ref{fig:S_rec_hist}b a logarithmic scale.
As shown in Figure \ref{fig:S_rec_hist}a, most of the spectra in the database are associated with stars that are not very active with the values of $S_{\rm rec}$ distributed around the mean $S_{\rm rec}$ value of the Sun ($\langle S_{\rm rec,\odot} \rangle = 0.223$; see Section \ref{sec:tri_vs_rec}). 
Figure \ref{fig:S_rec_hist}b demonstrates that there are a certain number of spectra having a higher value of $S_{\rm rec}$ which might be associated with active stars, 
and a dozen of spectra have isolated $S_{\rm rec}$ values greater than 1.1.
The features of the spectra with higher values of $S_{\rm rec}$ will be examined in detail in the further work.

\begin{figure}
    \epsscale{1.16}
	\plotone{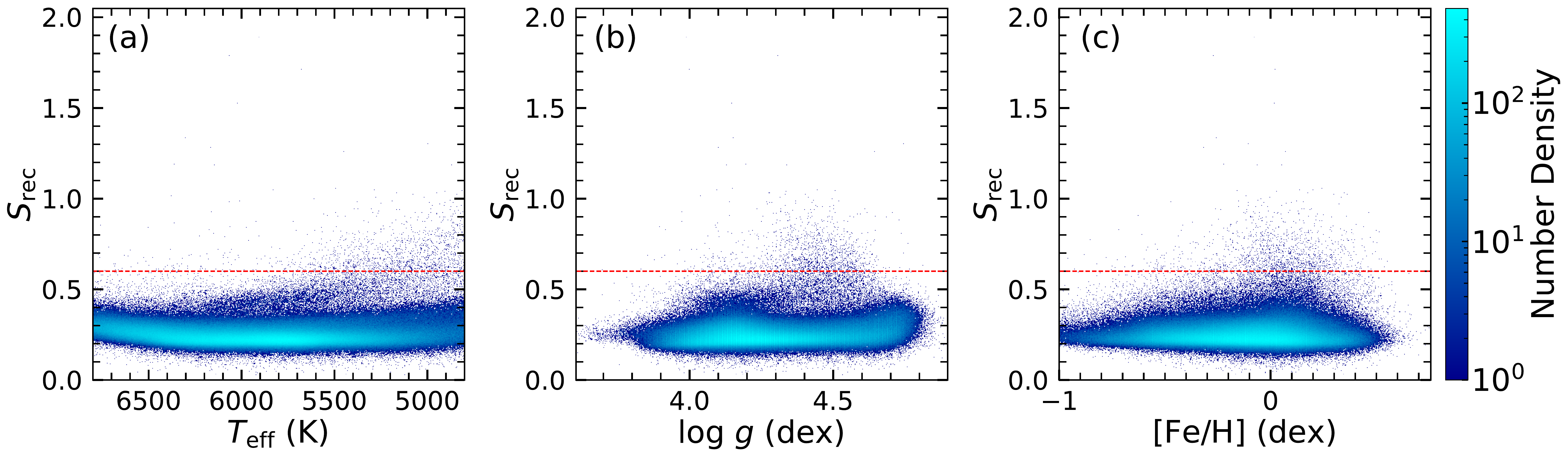}
	\caption{Scatter diagrams of (a) $T_{\rm eff}$ vs. $S_{\rm rec}$, (b) $\log\,g$ vs. $S_{\rm rec}$, and (c) [Fe/H] vs. $S_{\rm rec}$ for all the LAMOST LRS spectra of solar-like stars in the database.
	Color scale indicates number density.
	The horizontal dashed lines indicate $S_{\rm rec}=0.6$; 
	the sample above the line tends to appear in a compact range of $\log\,g$ (in panel b) and [Fe/H] (in panel c).
	} 
	\label{fig:S_rec-teff-logg-feh}
\end{figure}

The relationship between the stellar activity and the stellar parameters ($T_{\rm eff}$, $\log\,g$, and [Fe/H]) has attracted wide interest in the literature (e.g., \citealt{1968ApJ...153..221W, 2006AJ....132..161G, 2013A&A...549A.117M, 2013AJ....145..140Z, 2015RAA....15.1282Z, 2016A&A...595A..11L, 2018A&A...616A.108B, 2018MNRAS.476..908F, 2018ApJ...852...46K, 2019ApJ...887...84Z, 2021A&A...646A..77G}). 
In Figure \ref{fig:S_rec-teff-logg-feh}, we display the scatter diagrams of $T_{\rm eff}$ vs. $S_{\rm rec}$, $\log\,g$ vs. $S_{\rm rec}$, and [Fe/H] vs. $S_{\rm rec}$ using the values of the stellar parameters and activity indexes in the database, with color scale indicating number density.

\begin{figure}
    \epsscale{1.13}
	\plotone{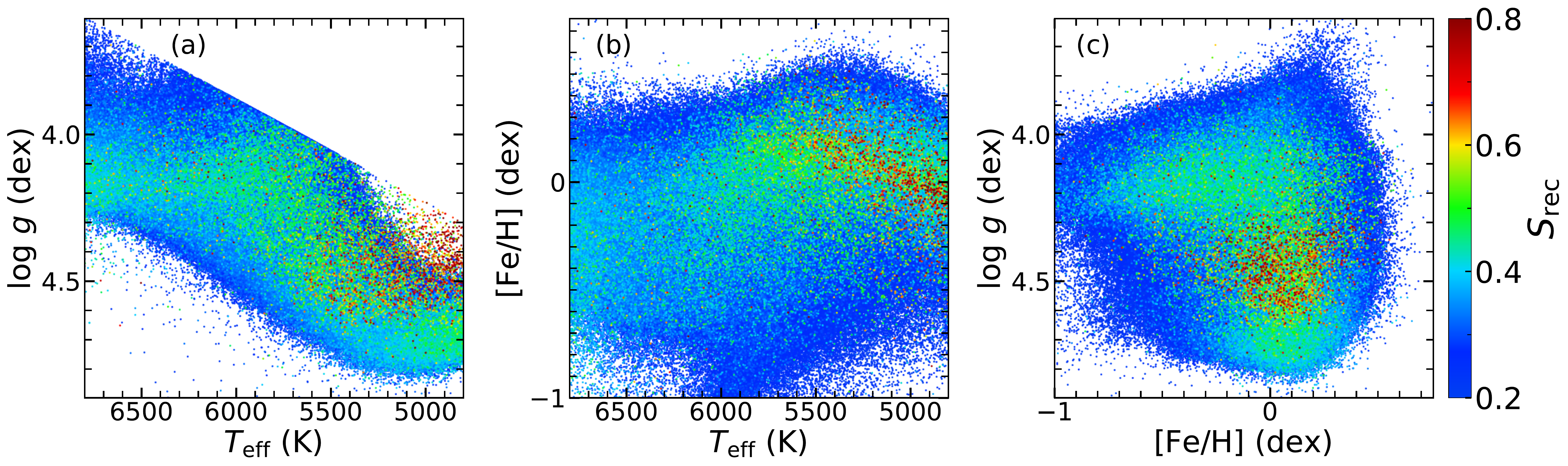}
	\caption{
	Distribution of $S_{\rm rec}$ values in scatter diagrams of (a) $T_{\rm eff}$ vs. $\log\,g$, (b) $T_{\rm eff}$ vs. [Fe/H], and (c) [Fe/H] vs. $\log\,g$ for all the LAMOST LRS spectra of solar-like stars in the database.
    The value of $S_{\rm rec}$ is indicated by color scale. 
    The data points with larger $S_{\rm rec}$ are overlaid on top of the data points with smaller $S_{\rm rec}$.
    }
	\label{fig:S_rec-teff-logg-feh_color}
\end{figure}

In Figure \ref{fig:S_rec-teff-logg-feh_color}, we display the distribution of $S_{\rm rec}$ values in scatter diagrams of $T_{\rm eff}$ vs. $\log\,g$, $T_{\rm eff}$ vs. [Fe/H], and [Fe/H] vs. $\log\,g$, with color scale indicating the value of $S_{\rm rec}$ (see the color bar in Figure \ref{fig:S_rec-teff-logg-feh_color}).
The smaller $S_{\rm rec}$ values ($<0.4$) are displayed in blue, the medium $S_{\rm rec}$ values (0.4--0.6) are green, and the larger $S_{\rm rec}$ values ($>0.6$) are red.
The data points in Figure \ref{fig:S_rec-teff-logg-feh_color} are drawn in order from smallest $S_{\rm rec}$ at the bottom to largest at the top, 
and hence the data points with larger $S_{\rm rec}$ values are overlaid on top of the data points with smaller $S_{\rm rec}$ values.
As shown in Figure \ref{fig:S_rec-teff-logg-feh_color}, most of the spectra have lower $S_{\rm rec}$ values (blue color). 
The higher the $S_{\rm rec}$ value, the smaller the distribution range is.

From the distribution diagrams of $S_{\rm rec}$ with respect to $\log\,g$ and [Fe/H] (Figures \ref{fig:S_rec-teff-logg-feh}b and c), 
it can be seen that the distribution morphology of the sample with $S_{\rm rec}>6.0$ is different from the sample with $S_{\rm rec}<6.0$. 
That is, the sample with $S_{\rm rec}>0.6$ tends to appear in a compact range of stellar parameters.
We draw a horizontal line indicating $S_{\rm rec}=0.6$ in Figures \ref{fig:S_rec-teff-logg-feh} to distinguish between the two regions with different morphologies of distribution.
The total number of the spectra with $S_{\rm rec}$ greater than 0.6 is 1424, which is about 0.1\% of the total spectra in the database.
Compared to previous investigations on the stellar chromospheric activity index distribution with respect to surface gravity and metallicity in the literature (e.g., \citealt{2006AJ....132..161G, 2008A&A...485..571J, 2011arXiv1107.5325L, 2013AJ....145..140Z, 2018ApJ...852...46K}), this feature of the $S_{\rm rec}$ distribution can be revealed thanks to the huge amounts of spectral data obtained by LAMOST.
With the assistance of Figure \ref{fig:S_rec-teff-logg-feh_color}, it can be determined that the stars with $S_{\rm rec}$ greater than 0.6 tend to appear in the parameter range of $T_{\rm eff}<5500\,{\rm K}$, $4.3<\log\,g<4.6$, and  $-0.2<{\rm [Fe/H]}<0.3$.

It should be noted that the $S_{\rm rec}$ as well as other $S$-index parameters (e.g., $S_{\rm tri}$ and $S_L$ in this work) is temperature (color) dependent by construction (e.g., \citealt{1968ApJ...153..221W}), as demonstrated by the curved baseline in Figure \ref{fig:S_rec-teff-logg-feh}a. The color-corrected indexes such as $R'_{\rm HK}$ \citep{1979ApJS...41...47L, 1984ApJ...279..763N} will be introduced for the LAMOST LRS spectra of solar-like stars in our future work.

\section{Summary and Conclusion}\label{sec:conclusion}
We obtain 1,330,654 high-quality LRS spectra of solar-like stars from LAMOST DR7.
The emission fluxes of Ca II H\&K line cores are measured by employing 1\,{\AA} rectangular bandpasses as well as 1.09\,{\AA} FWHM triangular bandpasses. 
The emission fluxes of pseudo-continuum on the two sides of the Ca II H\&K lines are measured by using 20\,{\AA} rectangular bandpasses.
We introduce three activity indexes $S_{\rm tri}$, $S_L$, and $S_{\rm rec}$ of Ca II H\&K lines, and calculate the values of the activity indexes for all the obtained LRS spectra of solar-like stars.
By examining the relations of the three indexes, we suggest that for the LAMOST LRS data, 
the $S_{\rm rec}$ index using rectangular bandpass is also a reliable indicator of chromospheric activity with a more straightforward physical meaning than the indexes using triangular bandpass.
Through the analysis of the $S_{\rm rec}$ distribution with stellar parameters, it is found that the solar-like stars with high level of chromospheric activity ($S_{\rm rec}>0.6$) tend to appear in the stellar parameter range of $T_{\rm eff}<5500\,{\rm K}$, $4.3<\log\,g<4.6$, and $-0.2<{\rm [Fe/H]}<0.3$.

A stellar chromospheric activity database of solar-like stars is constructed from all the 1,330,654 LAMOST LRS spectra obtained in this work.
The database is composed of a catalog of spectral sample and activity parameters and a library of spectrum diagrams of Ca II H\&K lines.
The catalog of the database provides the six derived emission flux measures of Ca II H\&K lines ($\widetilde{R}$, $\widetilde{V}$, $\widetilde{H}_{\rm tri}$, $\widetilde{K}_{\rm tri}$, $\widetilde{H}_{\rm rec}$, and $\widetilde{K}_{\rm rec}$), the three stellar chromospheric activity indexes ($S_{\rm tri}$, $S_L$, and $S_{\rm rec}$), and their uncertainties for all the LRS spectra employed by the database. 
The library of spectrum diagrams provide 1,330,654 spectral profile pictures of Ca II H\&K lines for the spectral sample in the database, in which the chromospheric activity parameters as well as the spectroscopic parameters are displayed.
All the components of the database are available online (see Section \ref{sec:database}).

Based on the database, one can investigate overall chromospheric activity properties of solar-like stars (such as the relation of activity index with stellar parameters, age, rotation period, etc.) and understand solar-stellar connection through chromospheric activity \citep{2021RNAAS...5....6H}.
We have shown the relationship between $S_{\rm rec}$ and stellar parameters $T_{\rm eff}$, $\log\,g$, and [Fe/H] in this work.
It is also important to investigate the connections of the stellar chromospheric activity revealed by the LAMOST spectral data with the stellar photospheric activity indicated by rotational modulation in the light curve data observed by Kepler/K2, TESS, etc. (e.g., \citealt{2015ApJS..221...18H, 2018ApJS..236....7H, 2017ApJ...834..207M, 2019NewA...66...31M, 2020Sci...368..518R, 2020ApJS..247....9Z, 2020ApJ...894L..11Z}), 
the stellar coronal activity revealed by the X-ray data observed by Chandra, XMM-Newton, etc. (e.g., \citealt{2019ApJ...871..193H,2020ApJ...902..114W}), 
and the stellar flare activity revealed by stellar light curve data and spectral data (e.g., \citealt{2012Natur.485..478M, 2018ApJS..236....7H, 2018MNRAS.479L.139L, 2019ApJS..244...37G, 2021ApJ...906...40G, 2021MNRAS.505L..79Y, 2022ApJ...928..180W}).
The time-domain data of LAMOST have been released (e.g., \citealt{2020arXiv200507210L, 2020ApJS..251...15Z, 2021RAA....21..249B, 2021RAA....21..292W}), and the method developed in this work can be used for time-domain analysis of stellar chromospheric activity. 
Stellar chromospheric activity information is also usefully for exploring habitability of exoplanet as well as exoplanet detection (e.g., \citealt{1997ApJ...485..319S, 2010ApJ...725..875I, 2011arXiv1107.5325L, 2021A&A...646A..77G}).

\section*{Acknowledgements}

This work is supported by the National Key R\&D Program of China (2019YFA0405000) and the National Natural Science Foundation of China (12073001 and 11973059). W.Z. and J.Z. thank the support of the Anhui Project (Z010118169). H.H. acknowledges the B-type Strategic Priority Program of the Chinese Academy of Sciences (XDB41000000), the CAS Strategic Pioneer Program on Space Science (XDA15052200), and the Astronomical Big Data Joint Research Center, co-founded by the National Astronomical Observatories, Chinese Academy of Sciences (NAOC) and the Alibaba Cloud. Guoshoujing Telescope (the Large Sky Area Multi-Object Fiber Spectroscopic Telescope, LAMOST) is a National Major Scientific Project built by the Chinese Academy of Sciences. Funding for the project has been provided by the National Development and Reform Commission. LAMOST is operated and managed by the National Astronomical Observatories, Chinese Academy of Sciences.

\vspace{5mm}

\facility{LAMOST}
\software{Astropy \citep{2013A&A...558A..33A, 2018AJ....156..123A},
          SciPy \citep{2020NatMe..17..261V},
          NumPy \citep{2007CSE.....9c..10O, 2011CSE....13b..22V,2020Natur.585..357H},
          Matplotlib \citep{2007CSE.....9...90H}
          }

\vspace{12mm}

\appendix

\section{Calibration Procedure for the Relationship Between {\texorpdfstring{$S_L$}{S\_L}} and {\texorpdfstring{$S_{\rm MWO}$}{S\_MWO}}} \label{sec:calibration_S_L_vs_S_MWO}

We identify 65 commons stars between the selected LRS spectra of solar-like stars in this work and the $S_{\rm MWO}$ catalog of MWO given by \citet{1991ApJS...76..383D}. 
We do not find common stars with the MWO catalog by \citet{1995ApJ...438..269B}. 
These common stars are listed in Table \ref{tab:common_stars_for_sindex_calibration} and used for $S_L$ and $S_{\rm MWO}$ relationship calibration.
The $S_{\rm MWO}$ values in Table \ref{tab:common_stars_for_sindex_calibration} are taken from \citet{1991ApJS...76..383D}, and
the $S_L$ values are taken from the catalog obtained by this work.
If a stellar object has more than one $S$-index records in a catalog, the median of the $S$-index records is adopted.

\begin{figure}
    \epsscale{0.5}
	\plotone{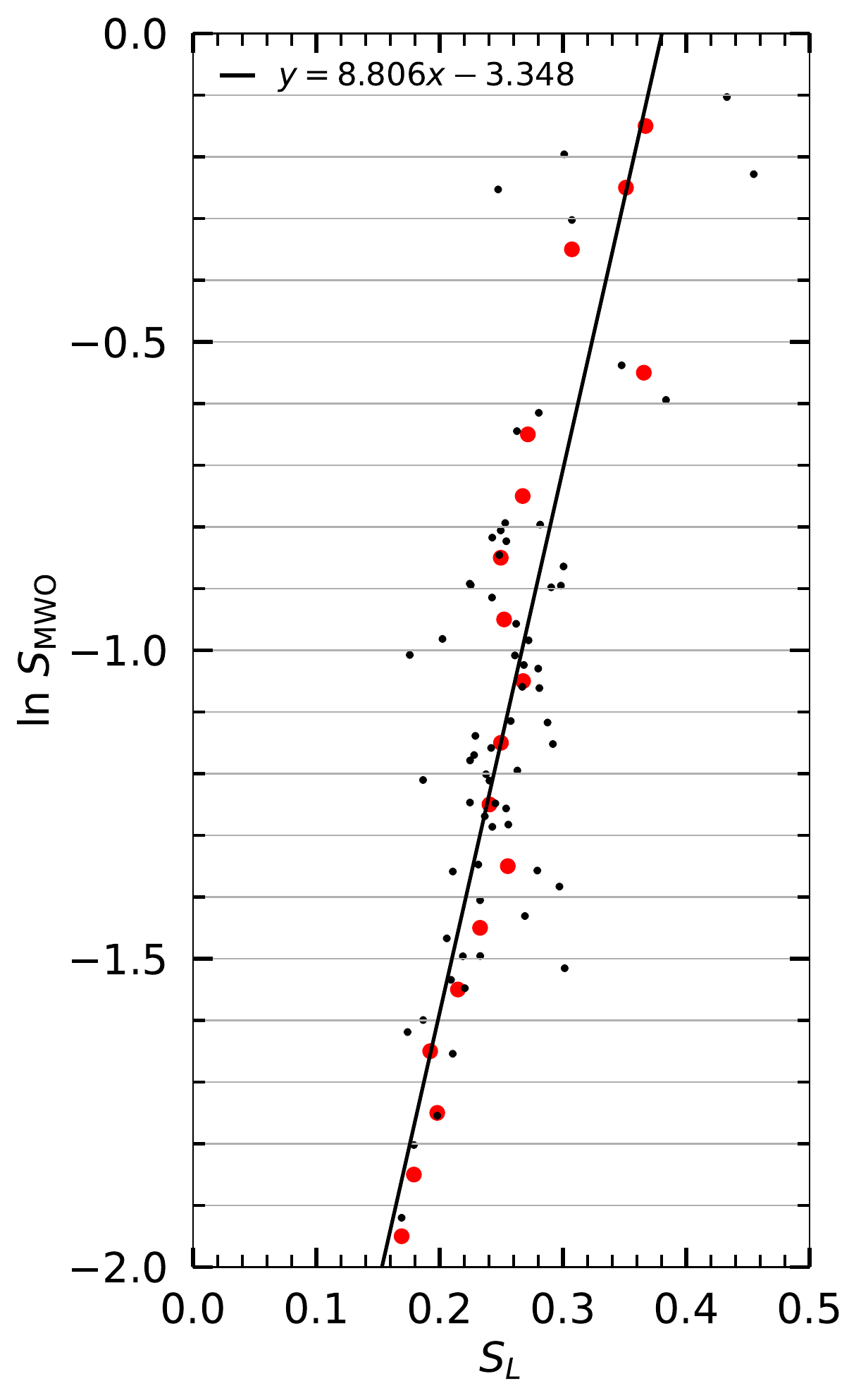}
	\caption{Linear fitting for the values of $S_L$ and $\ln S_{\rm MWO}$ of the common stars. Black dots are the original data as listed in Table \ref{tab:common_stars_for_sindex_calibration}. The values of $\ln S_{\rm MWO}$ of the common stars are distributed in the range from $-2$ to 0, which is divided into 20 equal bins (bin step = 0.1) distinguished by horizontal lines. The red dots indicate the median of the $S_L$ values in each bin. The linear fitting is performed for the red dots in the plot, and the fitting result is displayed as a black line. The formula for the fitted line is displayed, with $x$ representing $S_L$ and $y$ representing $\ln S_{\rm MWO}$.
	\label{fig:lnS_MWO-S_L}
	}
\end{figure}

To determine the parameters of the exponential relation between $S_L$ and $S_{\rm MWO}$ shown in Equation (\ref{eq:S_L_vs_S_MWO}), 
we perform a linear fitting for the values of $S_L$ and $\ln S_{\rm MWO}$ of the common stars, as illustrated in Figure \ref{fig:lnS_MWO-S_L}.
The original data of $S_L$ and $\ln S_{\rm MWO}$ of the common stars are displayed as black dots in Figure \ref{fig:lnS_MWO-S_L}.
It can be seen from Figure \ref{fig:lnS_MWO-S_L} that the values of $\ln S_{\rm MWO}$ of the common stars are not evenly distributed in the Y-axis direction.
To obtain a relatively evenly distributed data set, we divide the distribution range of the $\ln S_{\rm MWO}$ values (from $-2$ to 0) into 20 equal bins (bin step = 0.1) as illustrated in Figure \ref{fig:lnS_MWO-S_L},
and combine the data points in each bin into one data point which is displayed as a red dot in each bin in Figure \ref{fig:lnS_MWO-S_L}. 
The Y-coordinate of the combined data point in a bin is the center position of the bin in Y-axis, 
and the X-coordinate of the combined data point is the median of the $S_L$ values of the original data points within the bin.
The linear fitting is performed for the red dots in Figure \ref{fig:lnS_MWO-S_L}, and the fitting result is displayed as a black line. The formula for the fitted line is
\begin{equation} \label{eq:S_L_vs_lnS_MWO}
    \ln S_{\rm MWO} = 8.806\,S_L-3.348.
\end{equation}
Then, the exponential relation between $S_L$ and $S_{\rm MWO}$ in Equation (\ref{eq:S_L_vs_S_MWO}) can be derived from Equation (\ref{eq:S_L_vs_lnS_MWO}).

\startlongtable
\begin{deluxetable}{llllllllllll}
	\tablecaption{Common Stars Used for $S_L$ and $S_{\rm MWO}$ Relationship Calibration.
	\label{tab:common_stars_for_sindex_calibration}}
	\tablehead{
		& \colhead{Star Name} && \colhead{$S_L$} && \colhead{$\delta S_L$} && \colhead{$S_{\rm MWO}$} && \colhead{$\delta S_{\rm MWO}$} && \colhead{Gaia DR2 Source Identifier}
	}
	\startdata
& BD +58 1199 && 0.174 && 0.0089 && 0.198 && 0.0297 && Gaia DR2 1025549786973693312\\
& Cl Melotte 25 42 && 0.242 && 0.0033 && 0.314 && 0.0084 && Gaia DR2 145325548516513280\\
& HD 118576 && 0.179 && 0.0042 && 0.165 && 0.0030 && Gaia DR2 1456026937348507520\\
& HD 7983 && 0.187 && 0.0031 && 0.202 &&   && Gaia DR2 2471972451598042880\\
& HD 1342 && 0.198 && 0.0032 && 0.173 && 0.0033 && Gaia DR2 2547169391852546688\\
& Parenago 1199 && 0.176 && 0.0033 && 0.365 &&   && Gaia DR2 3017234389666905216\\
& Parenago 1626 && 0.348 && 0.0045 && 0.584 &&   && Gaia DR2 3017237653842191616\\
& Parenago 1158 && 0.202 && 0.0034 && 0.375 &&   && Gaia DR2 3209419916870340992\\
& Parenago 1361 && 0.225 && 0.0035 && 0.409 &&   && Gaia DR2 3209536636903447936\\
& Parenago 1241 && 0.290 && 0.0022 && 0.407 &&   && Gaia DR2 3209543989887451776\\
& Parenago 1322 && 0.433 && 0.0066 && 0.902 &&   && Gaia DR2 3209604390008665216\\
& Parenago 1357 && 0.186 && 0.0046 && 0.298 &&   && Gaia DR2 3209656071350169344\\
& BD +00 0873 && 0.169 && 0.0042 && 0.147 &&   && Gaia DR2 3231423476711449728\\
& Cl Melotte 25 99 && 0.280 && 0.0010 && 0.541 && 0.0273 && Gaia DR2 3309956850635519488\\
& Cl Melotte 25 46 && 0.269 && 0.0011 && 0.239 &&   && Gaia DR2 3311148828615843328\\
& Cl Melotte 25 25 && 0.297 && 0.0014 && 0.251 &&   && Gaia DR2 3312281669189449472\\
& Cl Melotte 25 79 && 0.250 && 0.0012 && 0.447 && 0.0353 && Gaia DR2 3314213025787054592\\
& Cl Melotte 111 150 && 0.384 && 0.0028 && 0.552 &&   && Gaia DR2 3960008298439044864\\
& Cl Melotte 111 92 && 0.238 && 0.0051 && 0.301 &&   && Gaia DR2 4008342623437661568\\
& Cl Melotte 111 97 && 0.240 && 0.0051 && 0.298 &&   && Gaia DR2 4008433608024885632\\
& Cl Melotte 111 85 && 0.225 && 0.0039 && 0.287 &&   && Gaia DR2 4008706733584941312\\
& Cl Melotte 111 132 && 0.245 && 0.0065 && 0.287 &&   && Gaia DR2 4008867674599508992\\
& Cl Melotte 111 118 && 0.206 && 0.0042 && 0.231 &&   && Gaia DR2 4009051048227419520\\
& Cl Melotte 111 86 && 0.233 && 0.0068 && 0.245 &&   && Gaia DR2 4009518100151157248\\
& Cl Melotte 25 7 && 0.301 && 0.0000 && 0.220 &&   && Gaia DR2 43538293935879680\\
& HD 149162 && 0.220 && 0.0046 && 0.213 && 0.0252 && Gaia DR2 4433380077474648192\\
& Cl Melotte 25 69 && 0.242 && 0.0019 && 0.401 && 0.0464 && Gaia DR2 48061409893621248\\
& Cl Melotte 25 43 && 0.280 && 0.0021 && 0.357 &&   && Gaia DR2 48197783694869760\\
& Cl Melotte 25 5 && 0.300 && 0.0047 && 0.421 && 0.0076 && Gaia DR2 64266768177592448\\
& Cl Melotte 22 2106 && 0.307 && 0.0049 && 0.739 &&   && Gaia DR2 64921458633614976\\
& Cl Melotte 22 2126 && 0.301 && 0.0058 && 0.822 &&   && Gaia DR2 64923279699744256\\
& Cl Melotte 22 2345 && 0.292 && 0.0040 && 0.316 && 0.0028 && Gaia DR2 64930495244783616\\
& Cl Melotte 22 1139 && 0.219 && 0.0035 && 0.224 &&   && Gaia DR2 64952829074688896\\
& Cl Melotte 22 923 && 0.253 && 0.0021 && 0.452 &&   && Gaia DR2 64971177174850304\\
& Cl Melotte 22 1797 && 0.267 && 0.0071 && 0.347 &&   && Gaia DR2 64999588381514496\\
& Cl Melotte 22 1215 && 0.248 && 0.0021 && 0.429 &&   && Gaia DR2 65004712279475712\\
& Cl Melotte 22 1613 && 0.225 && 0.0028 && 0.308 &&   && Gaia DR2 65020414679879296\\
& Cl Melotte 22 164 && 0.287 && 0.0070 && 0.327 &&   && Gaia DR2 65090680344356992\\
& Cl Melotte 22 1117 && 0.254 && 0.0021 && 0.285 &&   && Gaia DR2 65199978672758272\\
& Cl Melotte 22 708 && 0.281 && 0.0024 && 0.346 &&   && Gaia DR2 65222759179189248\\
& Cl Melotte 22 1122 && 0.237 && 0.0035 && 0.281 &&   && Gaia DR2 65225611037551360\\
& Cl Melotte 22 129 && 0.263 && 0.0033 && 0.525 &&   && Gaia DR2 65233788655261568\\
& Cl Melotte 22 233 && 0.211 && 0.0039 && 0.191 &&   && Gaia DR2 65242069352190976\\
& Cl Melotte 22 745 && 0.256 && 0.0032 && 0.277 &&   && Gaia DR2 65277975278721152\\
& Cl Melotte 22 489 && 0.272 && 0.0051 && 0.374 &&   && Gaia DR2 65295425728422528\\
& Cl* NGC 2632 KW 162 && 0.228 && 0.0032 && 0.310 &&   && Gaia DR2 661207024061875456\\
& Cl* NGC 2632 KW 301 && 0.224 && 0.0037 && 0.410 &&   && Gaia DR2 661212933936844416\\
& Cl* NGC 2632 KW 217 && 0.243 && 0.0050 && 0.276 &&   && Gaia DR2 661216743570426240\\
& Cl* NGC 2632 KW 250 && 0.209 && 0.0051 && 0.216 &&   && Gaia DR2 661298451027341056\\
& Cl* NGC 2632 KW 246 && 0.247 && 0.0092 && 0.776 &&   && Gaia DR2 661310923612365312\\
& Cl* NGC 2632 KW 238 && 0.261 && 0.0000 && 0.365 &&   && Gaia DR2 661311752544249088\\
& Cl* NGC 2632 KW 32 && 0.262 && 0.0000 && 0.384 &&   && Gaia DR2 664283079638402688\\
& Cl* NGC 2632 KW 58 && 0.243 && 0.0028 && 0.442 &&   && Gaia DR2 664311392062657920\\
& Cl* NGC 2632 KW 127 && 0.211 && 0.0027 && 0.257 &&   && Gaia DR2 664324684984105728\\
& Cl Melotte 22 2147 && 0.455 && 0.0035 && 0.796 &&   && Gaia DR2 66503449709270400\\
& Cl Melotte 22 1856 && 0.229 && 0.0025 && 0.320 &&   && Gaia DR2 66523893753437824\\
& Cl Melotte 22 2027 && 0.279 && 0.0042 && 0.257 &&   && Gaia DR2 66720946851771904\\
& Cl Melotte 22 1309 && 0.268 && 0.0034 && 0.359 &&   && Gaia DR2 66729261908482048\\
& Cl Melotte 22 996 && 0.254 && 0.0026 && 0.439 && 0.0120 && Gaia DR2 66788291938818304\\
& Cl Melotte 22 727 && 0.298 && 0.0072 && 0.408 &&   && Gaia DR2 66802654309459712\\
& Cl Melotte 22 1207 && 0.281 && 0.0029 && 0.451 &&   && Gaia DR2 66809491897360896\\
& Cl Melotte 25 4 && 0.263 && 0.0000 && 0.303 && 0.0414 && Gaia DR2 68001499939741440\\
& Cl Melotte 22 25 && 0.233 && 0.0035 && 0.224 &&   && Gaia DR2 68310561489710336\\
& Cl Melotte 22 405 && 0.258 && 0.0024 && 0.328 && 0.0332 && Gaia DR2 69811948914407168\\
& Cl Melotte 22 605 && 0.231 && 0.0049 && 0.260 &&   && Gaia DR2 69819404977607168\\
	\enddata
    \tablecomments{The $S_{\rm MWO}$ values of MWO are taken from \citet{1991ApJS...76..383D}. The star names are taken from the online catalog of the MWO data (\url{https://cdsarc.cds.unistra.fr/viz-bin/cat/III/159A}). The Gaia DR2 Source Identifiers are taken from the {\tt\string LAMOST LRS AFGK Catalog}. The $S_L$ values are taken from the catalog obtained by this work. If a stellar object has more than one $S$-index records in a catalog, the median of the $S$-index records is adopted. Some of the uncertainty values are blank because they are not available in the source catalog.
    }
\end{deluxetable}

\end{CJK*}
\end{document}